\documentclass{article}
\usepackage{amsmath}
\usepackage{amssymb}
\usepackage{amsfonts}
\usepackage{tikz}
\usepackage{color}
\usepackage{graphicx,caption}

\usetikzlibrary{decorations.pathmorphing}

\def\V{\mathcal V}

\newcounter{NN}
\newtheorem{theorem}[NN]{Theorem}

\begin{document}
\bibliographystyle{plain}
\title{Initial value problems for quad equations\\[1cm]} 
\author{Peter H.~van der Kamp}
\date{Department of Mathematics and Statistics\\
La Trobe University, Victoria 3086, Australia\\[5mm]
\today
}

\maketitle

\begin{abstract}
We describe a method to construct well-posed initial value problems for not necessarily integrable equations on not necessarily simply connected quad-graphs. Although the method does not always provide a well-posed initial value problem (not all quad-graphs admit well-posed initial value problems) it is effective in the class of rhombic embeddable quad-graphs. %The method is certainly not restricted to this class, as an example we
%present a well-posed initial value problem, for integrable equations, on the deltoidal trihexagonal tiling.
\end{abstract}

\section{Introduction}
By a {\em quad-graph} we mean a planar embedding (edges intersect only at their endpoints) of a connected undirected graph $G$ all of whose
faces (except possibly for the outer face if $G$ is finite) have degree 4. We refer to the faces of a quad-graph as {\em quads}. In a {\em non-simply
connected quad-graph} we allow certain faces to have degree different from 4, but greater than 1. These faces are called {\em holes}. The outer face
is also considered to be a hole and we do not exclude holes of degree 4. We say a quad-graph is {\em proper} if at any vertex there are at most two
edges bounding a single face. In the sequel we assume quad-graphs are proper quad-graphs. In general, $G$ need not be simple; we may have more
than one edge between distinct vertices, see Figure \ref{non-simple}. A non-proper non-simply connected quad-graph is shown in Figure \ref{non-proper}.

\parbox{55mm}{
\begin{center}
\begin{tikzpicture}[scale=.8]
\draw[thin]
    (1,0)--(.5,1)--(1,2)--(1,3)
    (1,0)--(1.5,1)--(1,2);
\draw (1,1.5) circle (1.5cm)
	;
\fill[white] (1,0) circle (0.09cm)
	(.5,1) circle (0.09cm)
	(1.5,1) circle (0.09cm)
	(1,2) circle (0.09cm)
	(1,3) circle (0.09cm)
	;
\draw (1,0) circle (0.09cm)
	(.5,1) circle (0.09cm)
	(1.5,1) circle (0.09cm)
	(1,2) circle (0.09cm)
	(1,3) circle (0.09cm)
	;
%\draw [thin] plot [smooth] coordinates {(1,0) (0,1) (1,3)};
%\draw [thin] plot [smooth] coordinates {(1,0) (2,1) (1,3)};
\end{tikzpicture}
\captionof{figure}{\label{non-simple} A non-simple quad-graph.}
\end{center}
} \hfill \parbox{58mm}{
\begin{center}
\begin{tikzpicture}[scale=1.183]
\draw[thin]
    (2,2)--(4,0)--(0,0)--(2,2);
\draw plot [smooth] coordinates { (2,2) (1.5,1) (2,.5) (2.5,1) (2,2) };
\fill[white] (0,0) circle (0.07cm)
	(2,2) circle (0.07cm)
	(4,0) circle (0.07cm)
	;
\draw (0,0) circle (0.07cm)
	(2,2) circle (0.07cm)
	(4,0) circle (0.07cm)
	;
%\draw [orange, thick] (2,-.25)--(2,.75) plot [smooth] coordinates {(0,1) (2,.25) (4,1) };
\end{tikzpicture}
\captionof{figure}{\label{non-proper} A non-proper quad-graph.}
\end{center}
}

%Note that a non-simply connected quad-graph is non-simple if it has a hole of degree 2.
A {\em strip} is a path of quads (each quad being adjacent along an edge to the previous quad) which does not turn:
on entering a face it exits across the opposite edge. We assume that the strips extend in both directions as far as possible.
Thus, the strips run from hole to hole (boundary to boundary), or to infinity if $G$ is infinite. Note, in Figure \ref{non-simple} there
is only one strip. We will represent the strips by orange dashed curves, as in Figure \ref{zero}.

%\smallskip
We assume a scalar field $u$ takes values on $\V(G)$, the set of vertices of $G$, and we impose relations between the values
of $u$ at the four vertices of each quad. There are no relations on  the holes. A {\em solution} is a set of values for $u$
such that at every quad of $G$ the imposed relation is satisfied. An assignment of values to a subset of $\V(G)$ is an
{\em initial value problem}. An initial value problem (IVP) is called, cf. \cite{IVP},
\begin{itemize}
\item[$\ast$]{\em Over-determined} if for generic initial values no solution exists,
\item[$\ast$]{\em Well-posed} if for generic initial values a solution exists and is unique,
\item[$\ast$]{\em Nearly well-posed} if for generic initial values a solution exist and $u$ takes at most finitely many values at each point,
\item[$\ast$]{\em Under-determined} if for generic initial values a solution exist, but additional values can be chosen freely.
\end{itemize}
%The third option is a special discrete feature, and does not relate to solutions in the continuous 
%An IVP is called {\em singular} if it is well-posed for generic initial values, but specific values are assigned in such a way
%that in the solution additional values can be chosen freely. A singular IVP yields a {\em singular solution}, see \cite[Definition 1]{ABS2}.

In this paper, we will take the relations to be given by (non-degenerate) multi-affine functions of degree 1 in each variable.
This ensures that given three values of $u$ on a quad the fourth value can be determined uniquely. Such equations play
an important role in the theory of discrete integrable systems. Here, parameters $\alpha$ are assigned to the strips, and the
relations on the quads are governed by one and the same equation which depends on the two parameters assigned to the strips
that cross at the quad. The equation has the right symmetry properties so that it doesn't matter
how the equation is oriented on any given quad. The key notion is {\em multi-dimensional consistency} which means that we can
consistently impose the equation on all faces of a multi-dimensional cube. A multi-dimensional consistent equation with the
right symmetry properties is called {\em integrable} for short.

%\smallskip
The problem of finding conditions for initial value problems for integrable equations on quad-graphs to be well-posed,
underdetermined, and overdetermined was considered by Adler and Veselov in \cite{AV}.  We quote their main result,
\cite[Theorem 2]{AV}:\footnote{Strips were called characteristics in \cite{AV}.}
\begin{quote} Let an equation be integrable [in the above sense], and let $P$ be a simple path in a finite simply connected planar quad-graph
$\Gamma$ without self-intersecting strips. Consider the Cauchy problem for this equation with generic initial data on the path $P$.
\begin{itemize}
\item[$\ast$] If each strip in $\Gamma$ intersects $P$ exactly at one edge then the IVP is well-posed, that is, the solution exists
and is unique.
\item[$\ast$] If some strip intersects $P$ more than once then the IVP is overdetermined (no solution exists for generic initial data).
\item[$\ast$] If some strip does not intersect $P$ then the IVP is underdetermined (if a solution exists, it is not unique).
\end{itemize}
\end{quote}
%\noindent
In \cite{AV} it was remarked that the assumption that $P$ is a simple path (a path with no repeated vertices) can be
relaxed and that the result is true for any connected subgraph. A connected subgraph $C$ is called a {\em Cauchy
subgraph} if it has the property that each strip meets $C$ exactly at one edge. Such a subgraph does not always
exists, which was illustrated by an example, \cite[Figure 17]{AV} (the deltoidal trihexagonal tiling). Two questions arose:
how to describe the set of general solutions on quad-graphs of such type, and whether an example without closed strips exists.

%\smallskip
In \cite{AV} it was implicitly assumed that solutions are non-singular, cf. \cite{Adler}, where this assumption was made explicit.
In section 2, we illustrate the theorem does not hold, in absence of this assumption. We present an IVP on a simple path which
intersects each strip exactly once but is nearly well-posed. This shows that the above condition for well-posedness is not sufficient.
However, for integrable equations, amongst the three solutions to this IVP, there is a unique non-singular solution. This is
the actual content of the theorem, as reformulated by Adler in his second thesis \cite{Adler}. We also present nearly well-posed IVPs
on graphs with a strip which is intersected twice and a strip which is not intersected by the path of initial values. These show that
the conditions for the IVP to be over- or under-determined are not sufficient. For integrable equations, all solutions to these IVPs
are singular, in accordance with the reformulated theorem in \cite{Adler}. The fact that the theorem only concerns IVPs on connected
subgraphs, has not been resolved in \cite{Adler}. We present a simple quad-graph without closed or self-intersecting strips which has
no Cauchy subgraph but does admit a well-posed IVP. This shows that specifying initial data on a path or Cauchy subgraph is too restrictive.
That the condition for well-posedness is not necessary is also shown by the IVP on the simpler graph given in Figure \ref{zero}.

\begin{center}
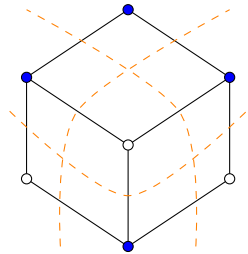

\begin{tikzpicture}[scale=.45]
\draw[thin]
    (4,0)--(1,2)--(1,5)--(4,7)--(7,5)--(7,2)--(4,0)--(4,3)--(7,5)
    (1,5)--(4,3);
\fill[white] (1,2) circle (0.15cm)
	(4,3) circle (0.15cm)
	(7,2) circle (0.15cm)
	;
\fill[blue] (1,5) circle (0.15cm)
	(7,5) circle (0.15cm)
	(4,0) circle (0.15cm)
	(4,7) circle (0.15cm)
	;
\draw (1,5) circle (0.15cm)
	(7,5) circle (0.15cm)
	(4,0) circle (0.15cm)
	(4,7) circle (0.15cm)
	(1,2) circle (0.15cm)
	(4,3) circle (0.15cm)
	(7,2) circle (0.15cm)
	;
\draw [orange, dashed] plot [smooth] coordinates {(2,0) (2.5,4) (7,7)};
\draw [orange, dashed] plot [smooth] coordinates {(6,0) (5.5,4) (1,7)};
\draw [orange, dashed] plot [smooth] coordinates {(0.5,4) (4,1.5) (7.5,4)};
\end{tikzpicture}
\captionof{figure}{\label{zero} A well-posed IVP (blue), which is not a Cauchy subgraph.}
\end{center}

%\smallskip
In section 3 we will show that whether an IVP is over-, or under-determined can be determined by counting the number of initial values. If there are too many the IVP is over-determined. If there are too few, the IVP is under-determined, or possibly over-determined if inconsistencies arise (i.e. there could be too many on a sub-graph). If the number of initial values is right, and no inconsistencies arise, the IVP is either well-posed, or nearly well-posed. Deciding on the well-posedness turns out to be more difficult. We will explain the difference between having generic relations on the quad (or a non-integrable equation) and imposing an integrable equation on the quads. In the generic case on can have self-intersecting strips but no closed strips. In the integrable case we can have closed strips (only contractable ones) but no self-intersecting strips.

In section 4 we prove that there exists no local condition for the well-posedness of an IVP in the general setting of quad-graphs.
The argument is simple; any well-posed IVP on a graph $G$ is nearly well-posed IVP on a graph $G^\prime$,
which is only slightly different from $G$ far away from the initial values.
In section 5 we present some examples of quad-graphs which do not admit well-posed IVPs.

In section 6 we present properties of strips of not necessarily simply-connected quad-graphs. We establish a one-to-one 
correspondence between intersecting curves in perforated planes satisfying these properties and connected proper quad-graphs.
Section 7 is devoted to constructing well-posed IVPs on quad-graphs. We show how the process of solving equations
defines a flow on the intersecting curves, which we represent by colouring, and we will show there are three obstructions
to well-posedness:
\begin{itemize}
\item[$\ast$]  Entanglement of curves might block the flow.
\item[$\ast$]  The flow out of a hole may clash with the flow in the graph.
\item[$\ast$] The flow may not encapsulate all holes.
\end{itemize}
Although our method is certainly not restricted to the setting of  rhombic embeddable quad-graphs,
%which are simply-connected quad-graphs whose strips cross each other at most once, cf. \cite{KS}, in that setting our 
here the method is guaranteed to produce a well-posed IVP.  In the final section we present a well-posed IVP for integrable equations
on the deltoidal trihexagonal tiling.

%\newpage
\section{Examples}
Figure  \ref{two} shows a path which intersects both strips exactly at one edge, cf. \cite[Example 2]{AV}.
According to \cite[Theorem 2]{AV} the IVP is well-posed. Instead, the IVP is nearly well-posed: in general the system of equations
$
f_1(u,u_1,v,u_{12})=f_2(u,v,w,u_{12})=f_3(u,w,u_2,u_{12})=0,
$
where $f_i$ is multi-affine, has three solutions.

\begin{center}
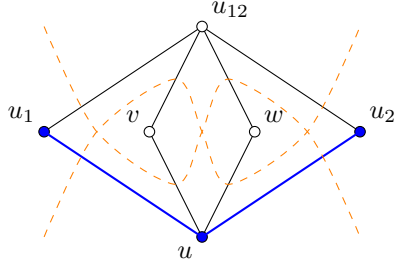

\begin{tikzpicture}[scale=1.4]
\draw[thin]
    (0,1)--(1.5,2)--(1,1)--(1.5,0)--(2,1)--(1.5,2)--(3,1);
\draw (1,1) node [anchor=south east] {$v$}
	(1.5,0) node [anchor=north east] {$u$}
	(0,1) node [anchor=south east] {$u_1$}
	(3,1) node [anchor=south west] {$u_2$}
	(2,1) node [anchor=south west] {$w$}
	(1.5,2) node [anchor=south west] {$u_{12}$}
	;
\fill[blue]
	(1.5,0) circle (0.05cm)
	(0,1) circle (0.05cm)
	(3,1) circle (0.05cm);
\fill[white]  (1,1) circle (0.05cm)
	(2,1) circle (0.05cm)
	(1.5,2) circle (0.05cm);
\draw (1,1) circle (0.05cm)
	(2,1) circle (0.05cm)
	(1.5,2) circle (0.05cm)
	(1.5,0) circle (0.05cm)
	(0,1) circle (0.05cm)
	(3,1) circle (0.05cm);
\draw[blue,thick]
    (0,1)--(1.5,0)--(3,1);
\draw [orange, dashed] plot [smooth] coordinates {(0,0) (.5,1) (1.25,1.5) (1.5,1) (1.75,.5) (2.5,1) (3,2)};
\draw [orange, dashed] plot [smooth] coordinates {(0,2) (.5,1) (1.25,.5) (1.5,1) (1.75,1.5) (2.5,1) (3,0)};
\end{tikzpicture}
\parbox{11.5cm}{\captionof{figure}{\label{two} The path $P$ (blue) intersects both strips exactly at one edge.}}
\end{center}
Figure \ref{one} shows a path which intersects one strip twice, whereas the other strip does not intersects the path, see  \cite[Example 1]{AV}.
According to \cite[Theorem 2]{AV} the IVP is both over-determined, and under-determined.
\begin{center}
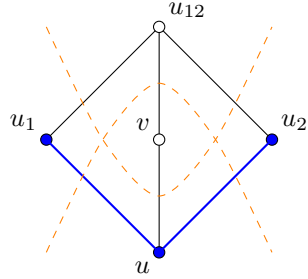

\begin{tikzpicture}[scale=1.5]
\draw (1,1) node [anchor=south east] {$v$}
	(1,2) node [anchor=south west] {$u_{12}$}
	(1,0) node [anchor=north east] {$u$}
	(0,1) node [anchor=south east] {$u_1$}
	(2,1) node [anchor=south west] {$u_2$};
\draw[thin]
(0,1)--(1,2)--(2,1)--(1,0)--(1,1)--(1,2);
\fill[blue]
	(1,0) circle (0.05cm)
	(0,1) circle (0.05cm)
	(2,1) circle (0.05cm);
\fill[white]
	(1,1) circle (0.05cm)
	(1,2) circle (0.05cm);		
\draw (1,1) circle (0.05cm)
	(1,2) circle (0.05cm)
	(1,0) circle (0.05cm)
	(0,1) circle (0.05cm)
	(2,1) circle (0.05cm);
\draw[blue,thick]
    (0,1)--(1,0)--(2,1);
\draw [orange, dashed] plot [smooth] coordinates {(0,0) (.5,1) (1,1.5) (1.5,1) (2,0)};
\draw [orange, dashed] plot [smooth] coordinates {(0,2) (.5,1) (1,.5) (1.5,1) (2,2)};
\end{tikzpicture}
\captionof{figure}{\label{one} A nearly well-posed initial value problem.}
\end{center}
However, the IVP is nearly well-posed, the number of solutions is easily determined, as follows.
Both multi-affine equations relate the value $u_{12}$ to the value $v$ through a
M\"obius transformation. Eliminating $u_{12}$ yields a quadratic equation for $v$ which has 2 solutions.
%\end{example}

The next example is also taken from \cite{AV}, where it was supposed to illustrate the difference
between integrable and non-integrable equations \cite[Example 5]{AV}.
Figure  \ref{three} shows a path which intersects one strip twice, another strip once and it does not intersect
the third strip.
\begin{center}
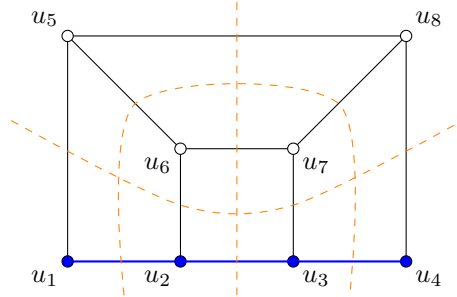

\begin{tikzpicture}[scale=1.5]
\draw 	(0,2) node [anchor=south east] {$u_5$}
	(1,1) node [anchor=north east] {$u_6$}
	(2,1) node [anchor=north west] {$u_7$}
	(3,2) node [anchor=south west] {$u_8$}
		(0,0) node [anchor=north east] {$u_1$}
		(1,0) node [anchor=north east] {$u_2$}
		(2,0) node [anchor=north west] {$u_3$}
		(3,0) node [anchor=north west] {$u_4$};
\draw[thin]
    (0,2)--(1,1)--(2,1)--(3,2)
    (0,0)--(0,2)--(3,2)--(3,0)
    (1,1)--(1,0)   (2,1)--(2,0);
\fill[white]
	(0,2) circle (0.05cm)
	(1,1) circle (0.05cm)
	(2,1) circle (0.05cm)
	(3,2) circle (0.05cm);
\fill[blue]
	(0,0) circle (0.05cm)
	(1,0) circle (0.05cm)
	(2,0) circle (0.05cm)
	(3,0) circle (0.05cm);
\draw
	(0,2) circle (0.05cm)
	(1,1) circle (0.05cm)
	(2,1) circle (0.05cm)
	(3,2) circle (0.05cm)
	(0,0) circle (0.05cm)
	(1,0) circle (0.05cm)
	(2,0) circle (0.05cm)
	(3,0) circle (0.05cm);
\draw[blue,thick]
    (0,0)--(3,0);
\draw [orange, dashed] plot [smooth] coordinates {(-.5,1.25) (1,.5) (2,.5) (3.5,1.25)};
\draw [orange, dashed] plot [smooth] coordinates {(1.5,-.3) (1.5,2.3)};
\draw [orange, dashed] plot [smooth] coordinates {(.5,-.3) (.6,1.4) (2.4,1.4) (2.5,-.3)};
\end{tikzpicture}
\captionof{figure}{\label{three} Another nearly well-posed IVP.}
\end{center}
\noindent
One can show (most easily using intersection theory) that the number of solutions to the system
\[
f_1(u_1,u_2,u_5,u_6)=f_2(u_2,u_3,u_6,u_7)=f_3(u_6,u_7,u_5,u_8)=f_4(u_3,u_4,u_7,u_8)=0,
\]
where $f_i$ is multi-affine, is four. This does not dependent on the integrability of the equation.

\subsection*{Some remarks on singular solutions (of integrable equations)}
A solution $s$ is {\em singular} (with respect to a vertex $v$) if at some quad,
with equation $E$, the equation $\partial_vE=0$ is also satisfied. We call a solution {\em regular} if it is not singular.
In \cite{JA} Atkinson introduced the notion of a {\em singular edge}, using certain bi-quadratics which can be associated to edges of quad-graphs with no self-intersecting strips in the setting of integrable equations. This he utilised to give a global description of admissible singularity configurations.

For integrable equations, most solutions to the nearly well-posed IVPs are singular.
The IVP in Figure \ref{two} has one regular solution and two singular solutions of the type depicted in
Figure \ref{se1}. The regular solution (as mentioned in \cite{AV}) satisfies the relation $E(u,u_1,u_2,u_{12},\alpha_1,\alpha_2)=0$,
but the singular solutions do not satisfy that relation. For the discrete potential KdV equation (or $H1$ in the classification of Adler,
Bobenko, and Suris \cite{ABS,ABS2}), we have
\[
E(u,u_1,u_2,u_{12},\alpha_1,\alpha_2)=(u-u_{12})(u_1-u_2)-\alpha_1+\alpha_2
\]
and the singular solution (of multiplicity 2) is $v=w=\infty,\ u_{12}=u$.

\parbox{55mm}{
\begin{center}
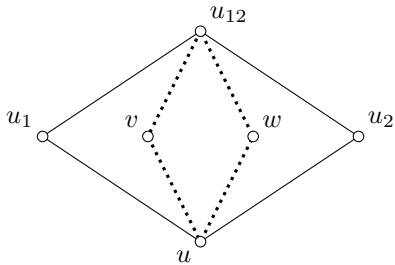

\begin{tikzpicture}[scale=1.4]
\draw (1,1) node [anchor=south east] {$v$}
	(1.5,0) node [anchor=north east] {$u$}
	(0,1) node [anchor=south east] {$u_1$}
	(3,1) node [anchor=south west] {$u_2$}
	(2,1) node [anchor=south west] {$w$}
	(1.5,2) node [anchor=south west] {$u_{12}$}
	;
\draw[thin]
    (0,1)--(1.5,0)--(3,1)--(1.5,2)--(0,1);
\draw[dotted, very thick]
	    (1.5,2)--(1,1)--(1.5,0)--(2,1)--(1.5,2);
\fill[white]
	(1.5,0) circle (0.05cm)
	(0,1) circle (0.05cm)
	(3,1) circle (0.05cm)
	(1,1) circle (0.05cm)
	(2,1) circle (0.05cm)
	(1.5,2) circle (0.05cm);
\draw (1,1) circle (0.05cm)
	(2,1) circle (0.05cm)
	(1.5,2) circle (0.05cm)
	(1.5,0) circle (0.05cm)
	(0,1) circle (0.05cm)
	(3,1) circle (0.05cm);
\end{tikzpicture}
\captionof{figure}{\label{se1} The singular edges are dotted.}
\end{center}
}
\hfill
\parbox{55mm}{
\begin{center}
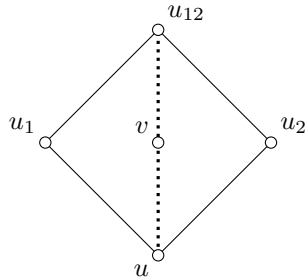

\begin{tikzpicture}[scale=1.5]
\draw (1,1) node [anchor=south east] {$v$}
	(1,2) node [anchor=south west] {$u_{12}$}
	(1,0) node [anchor=north east] {$u$}
	(0,1) node [anchor=south east] {$u_1$}
	(2,1) node [anchor=south west] {$u_2$};
\draw[thin]
	(0,1)--(1,2)--(2,1)--(1,0)--(0,1);
\draw[dotted, very thick]
	(1,0)--(1,2);
\fill[white]
	(1,1) circle (0.05cm)
	(1,2) circle (0.05cm)
	(1,0) circle (0.05cm)
	(0,1) circle (0.05cm)
	(2,1) circle (0.05cm);	
\draw (1,1) circle (0.05cm)
	(1,2) circle (0.05cm)
	(1,0) circle (0.05cm)
	(0,1) circle (0.05cm)
	(2,1) circle (0.05cm);
\end{tikzpicture}
\captionof{figure}{\label{se2} The singular edges for the solution to the IVP of Figure \ref{one}.}
\end{center}
}

Both solutions to the IVP in Figure \ref{one} are singular with singularity configuration given
in Figure \ref{se2}. For $H1$ the solution is $v=u_{12}=\infty$ (with multiplicity 2), but for other equations of the
Adler, Bobenko, and Suris list \cite{ABS}, the solutions $\{v,u_{12}\}$ are finite and depend on $u$ only.

In Figure \ref{three} one can {\em not} consider the quads as the faces of a 3d cube, since
$\{u_1,u_2,u_3,u_4\}$ is generic initial data. For integrable equations each solution is singular, with
one of the two singularity configurations given in Figure \ref{se3}. The solution $\{u_5,u_6,u_7,u_8\}$
does not depend on either $u_1$ or $u_4$.

All these observations are in accordance with \cite[Theorem 3.10]{Adler} which states that an IVP for
integrable equations on a path $P$ which intersects every strip exactly once has a unique regular
solution, that if a strip is intersected more than once no regular solutions exists, and if a strip is
not intersected by $P$ the existence of a regular solution implies non-uniqueness.

\begin{center}
\parbox{57mm}{
\begin{center}
\begin{tikzpicture}[scale=1.2]
\draw 	(0,2) node [anchor=south east] {$u_5$}
	(1,1) node [anchor=north east] {$u_6$}
	(2,1) node [anchor=north west] {$u_7$}
	(3,2) node [anchor=south west] {$u_8$}
		(0,0) node [anchor=north east] {$u_1$}
		(1,0) node [anchor=north east] {$u_2$}
		(2,0) node [anchor=north west] {$u_3$}
		(3,0) node [anchor=north west] {$u_4$};
\draw[thin]
    (0,0)--(0,2)--(3,2)--(3,0)--(0,0)
    (3,2)--(2,1)--(2,0);
\draw[dotted, very thick]
    (1,0)--(1,1)--(0,2)
    (1,1)--(2,1);
\fill[white]
	(0,2) circle (0.05cm)
	(1,1) circle (0.05cm)
	(2,1) circle (0.05cm)
	(3,2) circle (0.05cm)
	(0,0) circle (0.05cm)
	(1,0) circle (0.05cm)
	(2,0) circle (0.05cm)
	(3,0) circle (0.05cm);
\draw
	(0,2) circle (0.05cm)
	(1,1) circle (0.05cm)
	(2,1) circle (0.05cm)
	(3,2) circle (0.05cm)
	(0,0) circle (0.05cm)
	(1,0) circle (0.05cm)
	(2,0) circle (0.05cm)
	(3,0) circle (0.05cm);
\end{tikzpicture}
\end{center}}
\hfill
\parbox{57mm}{
\begin{center}
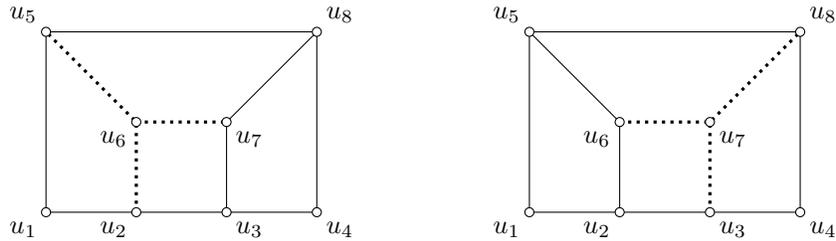

\begin{tikzpicture}[scale=1.2]
\draw 	(0,2) node [anchor=south east] {$u_5$}
	(1,1) node [anchor=north east] {$u_6$}
	(2,1) node [anchor=north west] {$u_7$}
	(3,2) node [anchor=south west] {$u_8$}
		(0,0) node [anchor=north east] {$u_1$}
		(1,0) node [anchor=north east] {$u_2$}
		(2,0) node [anchor=north west] {$u_3$}
		(3,0) node [anchor=north west] {$u_4$};
\draw[thin]
    (0,0)--(0,2)--(3,2)--(3,0)--(0,0)
    (0,2)--(1,1)--(1,0);
\draw[dotted, very thick]
    (2,0)--(2,1)--(3,2)
    (1,1)--(2,1);
\fill[white]
	(0,2) circle (0.05cm)
	(1,1) circle (0.05cm)
	(2,1) circle (0.05cm)
	(3,2) circle (0.05cm)
	(0,0) circle (0.05cm)
	(1,0) circle (0.05cm)
	(2,0) circle (0.05cm)
	(3,0) circle (0.05cm);
\draw
	(0,2) circle (0.05cm)
	(1,1) circle (0.05cm)
	(2,1) circle (0.05cm)
	(3,2) circle (0.05cm)
	(0,0) circle (0.05cm)
	(1,0) circle (0.05cm)
	(2,0) circle (0.05cm)
	(3,0) circle (0.05cm);
\end{tikzpicture}
\end{center}}
\parbox{11.5cm}{\captionof{figure}{\label{se3} Singularity configurations of the solution to the IVP in Figure \ref{three}.}
}
\end{center}

The last example in this section is a quad-graph without closed strips which has no Cauchy subgraph.
\begin{center}
\begin{tikzpicture}[scale=.48]
\draw[thin]
    (1,4)--(2,6)--(2,8)--(1,10)--(3,11)--(4,13)--(6,12)--(8,12)--(10,13)--(11,11)--(13,10)--(12,8)--(12,6)--(13,4)--(11,3)--(10,1)--(8,2)--(6,2)--(4,1)--(3,3)--(1,4)
    (2,6)--(5,5)--(6,2)
    (5,5)--(5,9)
    (8,12)--(9,9)
    (12,6)--(9,5)
    (5.125,5)--(9,5);
\fill[white] (1,4) circle (0.125cm)
	(1,10) circle (0.125cm)
	(2,6) circle (0.125cm)
	(4,1) circle (0.125cm)
	(4,13) circle (0.125cm)
	(6,2) circle (0.125cm)
	(8,12) circle (0.125cm)
	(10,1) circle (0.125cm)
	(10,13) circle (0.125cm)
	(12,6) circle (0.125cm)
	(13,4) circle (0.125cm)
	(13,10) circle (0.125cm)
	(5,5) circle (0.125cm);
\fill[blue]
	(2,8) circle (0.125cm)
	(3,11) circle (0.125cm)
	(5,9) circle (0.125cm)
	(6,12) circle (0.125cm)
	(9,9) circle (0.125cm)
	(11,11) circle (0.125cm)
	(12,8) circle (0.125cm)
	(11,3) circle (0.125cm)
	(9,5) circle (0.125cm)
	(8,2) circle (0.125cm)
	(3,3) circle (0.125cm);
\draw (1,4) circle (0.125cm)
	(1,10) circle (0.125cm)
	(2,6) circle (0.125cm)
	(4,1) circle (0.125cm)
	(4,13) circle (0.125cm)
	(6,2) circle (0.125cm)
	(8,12) circle (0.125cm)
	(10,1) circle (0.125cm)
	(10,13) circle (0.125cm)
	(12,6) circle (0.125cm)
	(13,4) circle (0.125cm)
	(13,10) circle (0.125cm)
	(2,8) circle (0.125cm)
	(3,11) circle (0.125cm)
	(5,9) circle (0.125cm)
	(6,12) circle (0.125cm)
	(9,9) circle (0.125cm)
	(11,11) circle (0.125cm)
	(12,8) circle (0.125cm)
	(11,3) circle (0.125cm)
	(9,5) circle (0.125cm)
	(8,2) circle (0.125cm)
	(3,3) circle (0.125cm)
	(5,5) circle (0.125cm);
\draw[color=blue,thick]
    (2,8)--(5,9) (3,11)--(5,9)--(9,9)--(11,11) (12,8)--(9,9)--(9,5)--(11,3) (8,2)--(9,5) (3,3)--(4.9,4.9) (5,9)--(6,12);
%\draw[color=blue, thick, style=dashed]
\draw [orange, dashed] plot [smooth] coordinates {(1,2) (4,7) (1,12)};
\draw [orange, dashed] plot [smooth] coordinates {(2,13)(7,10)(12,13)};
\draw [orange, dashed] plot [smooth] coordinates {(13,12)(10,7)(13,2)};
\draw [orange, dashed] plot [smooth] coordinates {(12,1)(7,4)(2,1)};
\draw [orange, dashed] plot [smooth] coordinates {(0,5)(4,4)(5,0)};
\draw [orange, dashed] plot [smooth] coordinates {(0,9)(4,10)(5,14)};
\draw [orange, dashed] plot [smooth] coordinates {(9,14)(10,10)(14,9)};
\draw [orange, dashed] plot [smooth] coordinates {(14,5)(10,4)(9,0)};
\draw [orange, dashed] plot [smooth] coordinates {(0,7)(14,7)};
\draw [orange, dashed] plot [smooth] coordinates {(7,0)(7,14)};
\end{tikzpicture}
\parbox{11cm}{
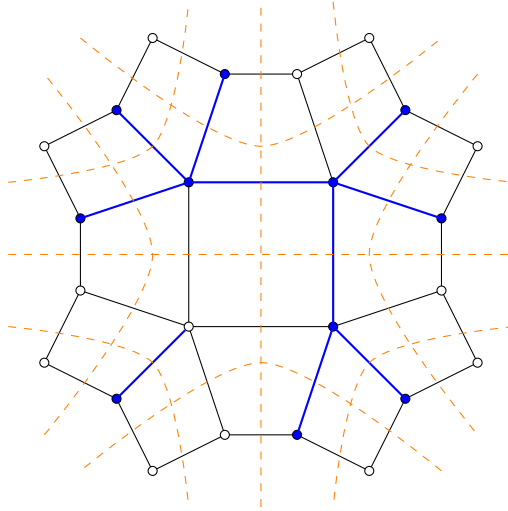
\captionof{figure}{\label{four} A well-posed IVP on a quad-graph without closed strips which has no Cauchy subgraph.}}
\end{center}
In Figure \ref{four} the disconnected blue subgraph intersects every strip exactly once, but if we were to connect the two components
the subgraph would intersect the vertical or horizontal strip twice. The IVP given by the collection of blue dots is well-posed
(note that on the blue subgraph one vertex is not part of the IVP). 

\section{The right number of initial values}
Consider a quad-graph, e.g. the graph in Figure \ref{hole}. Let $q$ be the number of quads, $v$ the number of vertices, $s$ the number of strips, and $c$ the number of closed strips. We let $h$ count the number of holes excluding the outer one, where we assume, without loss of generality, that holes do not share edges.\footnote{Without this assumption $h$ should count the number of connected sets of holes, where two holes are connected if they share an edge. Because quad-graphs are connected, any connected set of holes is simply
connected. By removing all shared edges any connected set of holes turns into a single (cycle graph) hole.} 
\begin{center}
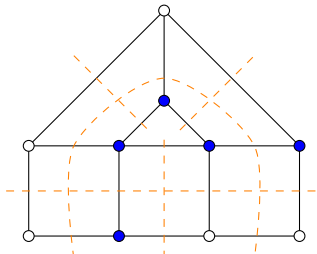

\begin{tikzpicture}[scale=.6]
\draw[thin]
    (0,0)--(0,2)--(3,5)--(6,2)--(6,0)--(0,0)
    (0,2)--(6,2)
    (2,0)--(2,2)--(3,3)--(4,2)--(4,0)
    (3,3)--(3,5);
\fill[white] (0,0) circle (0.12cm)
	(4,0) circle (0.12cm)
	(6,0) circle (0.12cm)
	(0,2) circle (0.12cm)
	(3,5) circle (0.12cm);
\fill[blue]	(2,0) circle (0.12cm)
	(2,2) circle (0.12cm)
	(3,3) circle (0.12cm)
	(4,2) circle (0.12cm)
	(6,2) circle (0.12cm)
	;
\draw (0,0) circle (0.12cm)
	(4,0) circle (0.12cm)
	(6,0) circle (0.12cm)
	(0,2) circle (0.12cm)
	(3,5) circle (0.12cm)
	(2,0) circle (0.12cm)
	(2,2) circle (0.12cm)
	(3,3) circle (0.12cm)
	(4,2) circle (0.12cm)
	(6,2) circle (0.12cm)
	;
\draw [orange, dashed]  plot [smooth] coordinates {(1,-.5) (1,2) (3,3.5) (5,2) (5,-.5)};
\draw [orange, dashed]  (-.5,1)--(6.5,1)
	(3,-.5)--(3,2.25)
	(1,4)--(2.65,2.35)
	(3.35,2.35)--(5,4);
\end{tikzpicture}
\parbox{11.5cm}{\captionof{figure}{\label{hole} A well-posed IVP on a non-simply connected quad-graph. Here $v=10$, $h=1$, $c=0$, $s=5$, and $q=5$.}}
\end{center}
We have the following relation:
\begin{equation} \label{inv}
v+h+c=s+q+1.
\end{equation}
This can be shown by using first of all Euler's formula which states that for any planar graph $v-e+f=1$ (where $f=q+h$ is the number of faces excluding the outer one and $e$ is the number of edges), secondly the fact that $4q=2e_i+e_b$ (where $e_i$ and $e_b$ are the internal edges and the edges on the boundary respectively), and thirdly by noting that every non-closed strip crosses the boundary twice, i.e. $e_b=2(s-c)$.
% by induction on the number of quads and holes in the graph.
 %For example, by adding a quad $Q$ to a graph $G$ (either to the
%outside hole or to an inner one), such that $Q$ has one edge in common with $G$, we raise $q$ by 1, $s$ by 1, and $v$ by 2, which is a symmetry
%of equation (\ref{inv}). Holes are created by removing quads that are not on the boundary. Another way to prove relation (\ref{inv}) is by summing
%angles, cf. \cite{AV}. However, this {\em only} works for simple graphs, as F\'ary's theorem states that any {\em simple} planar graph can be drawn without crossings so that its edges are straight line segments. %Then we may assume that all faces of $G$ are polygons, and use the simple fact that the sum
%of angles in a polygon $P$ with $p$ edges is $(p-2)\pi$. Let $S$ denote the sum of angles in the quads of $G$. Obviously, we have $S=2\pi q$.
%We find another expression for $S$ by taking the sum over the vertices. Let $b_0$ be the number of vertices that belong to the outer boundary,
%$b_i$ ($1\leq i\leq h$) be the number of vertices of hole $i$, and set $b=\sum_{i=0}^h b_i$. From the inner vertices we get a contribution $2\pi(v-b)$.
%The vertices on the outer boundary yield $(b_0-2)\pi$, and the angles around hole $i$ add up to $2\pi b_i-(b_i-2)\pi=\pi b_i + 2 \pi$. In total we find
%$S=2\pi(v+\frac{b}{2}+h-1)$. The result (\ref{inv}) follows by noting that every non-closed strip crosses the boundary twice, that is, we have
%$b=2(s-c)$.

\subsection{Generic relations (non-integrable equations)} \label{rinu}
For generic relations on the quads of a graph $G$ the right number of initial values is determined as follows. There are $v$ variables and
$q$ equations, so we need to specify $v-q$ values initially. Using relation (\ref{inv})  the right number of initial values $n$ equals
\[
n=s+1-h-c.
\]
Then, if the set of (generic, multi-affine) equations is triangular\footnote{A set of $q$ polynomials $\{p_1,p_2,\ldots,p_q\}$ in $q$ variables (we consider the initial values as parameters here) is triangular if, for some ordering of the variables and polynomials, the polynomial $p_i$ depends on the first $i$ variables only, for all $i$.} every equation determines the value of one variable uniquely and the IVP is well-posed. If the set of equations is not triangular there are multiple solutions, the IVP is nearly well-posed. In general, it is not easy to determine whether a set of polynomials is triangular, let alone to determine which variables to choose as parameters so that the remaining system is triangular.

One thing that {\em is} clear is that if a graph contains a closed strip, $c\neq0$, no matter which variables we chose as parameters, the remaining
set of polynomials will not be triangular, see Figures \ref{closed1} and \ref{closed2}.

\parbox{57mm}{
\begin{center}
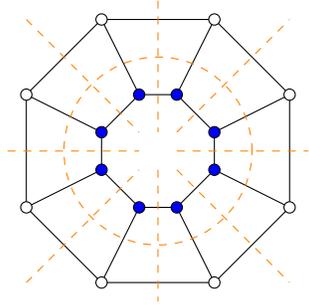

\begin{tikzpicture}[scale=.5]
\draw[thin]
    (0,2)--(0,5)--(2,7)--(5,7)--(7,5)--(7,2)--(5,0)--(2,0)--(0,2)--(2,3)--(2,4)--(3,5)--(4,5)--(5,4)--(5,3)--(4,2)--(3,2)--(2,3)
    (0,5)--(2,4)
    (2,7)--(3,5)
    (5,7)--(4,5)
    (5,4)--(7,5)
    (5,3)--(7,2)
    (5,0)--(4,2)
    (2,0)--(3,2);
\fill[white] (0,2) circle (0.15cm)
	(0,5) circle (0.15cm)
	(2,0) circle (0.15cm)
	(2,7) circle (0.15cm)
	(5,0) circle (0.15cm)
	(5,7) circle (0.15cm)
	(7,2) circle (0.15cm)
	(7,5) circle (0.15cm)
	;
\fill[blue]	(2,3) circle (0.15cm)
	(2,4) circle (0.15cm)
	(3,2) circle (0.15cm)
	(3,5) circle (0.15cm)
	(4,2) circle (0.15cm)
	(4,5) circle (0.15cm)
	(5,3) circle (0.15cm)
	(5,4) circle (0.15cm)
	;
\draw (0,2) circle (0.15cm)
	(0,5) circle (0.15cm)
	(2,0) circle (0.15cm)
	(2,7) circle (0.15cm)
	(5,0) circle (0.15cm)
	(5,7) circle (0.15cm)
	(7,2) circle (0.15cm)
	(7,5) circle (0.15cm)
	(2,3) circle (0.15cm)
	(2,4) circle (0.15cm)
	(3,2) circle (0.15cm)
	(3,5) circle (0.15cm)
	(4,2) circle (0.15cm)
	(4,5) circle (0.15cm)
	(5,3) circle (0.15cm)
	(5,4) circle (0.15cm)
	;
\draw [orange, dashed]  (3.5,3.5) circle (2.5cm);
\draw [orange, dashed]  (-.5,3.5)--(3,3.5)
	(0,7)--(3,4)
	(3.5,7.5)--(3.5,4)
	(4,4)--(7,7)
	(4,3.5)--(7.5,3.5)
	(4,3)--(7,0)
	(3.5,3)--(3.5,-.5)
	(0,0)--(3,3);
\end{tikzpicture}
\captionof{figure}{\label{closed1} For these initial values the number of solutions is minimal: 2.}
\end{center}
} \hfill \parbox{57mm}{
\begin{center}
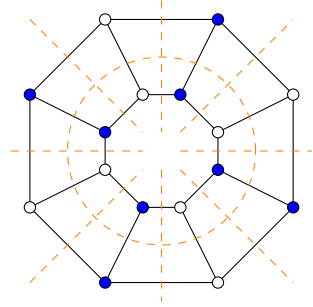

\begin{tikzpicture}[scale=.5]
\draw[thin]
    (0,2)--(0,5)--(2,7)--(5,7)--(7,5)--(7,2)--(5,0)--(2,0)--(0,2)--(2,3)--(2,4)--(3,5)--(4,5)--(5,4)--(5,3)--(4,2)--(3,2)--(2,3)
    (0,5)--(2,4)
    (2,7)--(3,5)
    (5,7)--(4,5)
    (5,4)--(7,5)
    (5,3)--(7,2)
    (5,0)--(4,2)
    (2,0)--(3,2);
\fill[white] (0,2) circle (0.15cm)
	(2,3) circle (0.15cm)
	(2,7) circle (0.15cm)
	(3,5) circle (0.15cm)
	(4,2) circle (0.15cm)
	(5,0) circle (0.15cm)
	(5,4) circle (0.15cm)
	(7,5) circle (0.15cm)
	;
\fill[blue]
	(4,5) circle (0.15cm)
	(5,7) circle (0.15cm)
	(5,3) circle (0.15cm)
	(7,2) circle (0.15cm)
	(2,0) circle (0.15cm)
	(3,2) circle (0.15cm)
	(0,5) circle (0.15cm)
	(2,4) circle (0.15cm)
	;
\draw (0,2) circle (0.15cm)
	(2,3) circle (0.15cm)
	(2,7) circle (0.15cm)
	(3,5) circle (0.15cm)
	(4,2) circle (0.15cm)
	(5,0) circle (0.15cm)
	(5,4) circle (0.15cm)
	(7,5) circle (0.15cm)
	(4,5) circle (0.15cm)
	(5,7) circle (0.15cm)
	(5,3) circle (0.15cm)
	(7,2) circle (0.15cm)
	(2,0) circle (0.15cm)
	(3,2) circle (0.15cm)
	(0,5) circle (0.15cm)
	(2,4) circle (0.15cm)
	;
\draw [orange, dashed]  (3.5,3.5) circle (2.5cm);
\draw [orange, dashed]  (-.5,3.5)--(3,3.5)
	(0,7)--(3,4)
	(3.5,7.5)--(3.5,4)
	(4,4)--(7,7)
	(4,3.5)--(7.5,3.5)
	(4,3)--(7,0)
	(3.5,3)--(3.5,-.5)
	(0,0)--(3,3);
\end{tikzpicture}
\captionof{figure}{\label{closed2} For these initial values the number of solutions is maximal: $2^4$.}
\end{center}
}

Also clear are the following conditions. Let $n$ denote the number of initial values assigned by an IVP. If $n>s+1-h-c$ the IVP
is over-determined. If $n<s+1-h-c$ the IVP is either under-determined (the solution contains arbitrary parameters) or over-determined
(the system is inconsistent if there is a sub-graph $G^\prime$ for which $n^\prime>s^\prime+1-h^\prime-c^\prime$). If $n=s+1-h-c$, and
the system is not inconsistent (we have $n^\prime\leq s^\prime+1-h^\prime-c^\prime$ for all sub-graphs $G^\prime$), the IVP is either 
ell-posed or nearly well-posed.

\subsection{Integrable equations}
Let us consider the simplest quad-graph with a closed strip, as in Figure  \ref{seven}.

\begin{center}
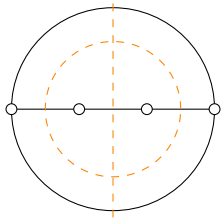

\begin{tikzpicture}[scale=.9]
\draw    (0,1)--(3,1)
	(1.5,1) circle (1.5cm)
	;
\fill[white] (3,1) circle (0.08cm)
	(0,1) circle (0.08cm)
	(1,1) circle (0.08cm)
	(2,1) circle (0.08cm)
	;
\draw (3,1) circle (0.08cm)
	(0,1) circle (0.08cm)
	(1,1) circle (0.08cm)
	(2,1) circle (0.08cm)
	;
\draw [orange, dashed]  (1.5,1) circle (1cm);
\draw [orange, dashed]  (1.5,-.6)--(1.5,2.6);
%\draw plot [smooth] coordinates {(0,2) (2.2,2) (3,0)};
%\draw plot [smooth] coordinates {(0,2) (.8,0) (3,0)};
\end{tikzpicture}
\captionof{figure}{\label{seven} Two relations on the same four vertices.}
\end{center}

As we saw in the previous section, for generic relations on its two quads no well-posed IVP exists. Any set of two vertices will provide a
nearly well-posed IVP. However, if we assume the {\em same} equation on both quads then any set of three initial values provides a
well-posed IVP.

\begin{center}
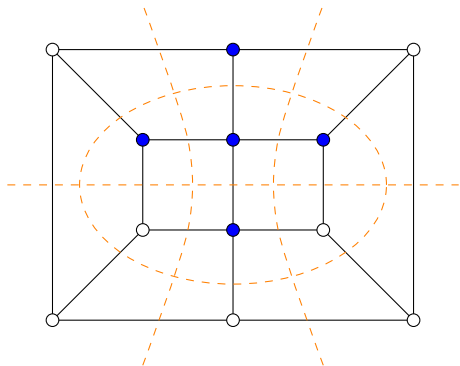

\begin{tikzpicture}[scale=1.2]
\draw[thin]
    (0,0)--(0,3)--(4,3)--(4,0)--(0,0)--(1,1)--(3,1)--(4,0)
    (2,0)--(2,3)
    (0,3)--(1,2)--(1,1)
    (1,2)--(3,2)
    (4,3)--(3,2)--(3,1);
\fill[white] (0,0) circle (0.07cm)
	(0,3) circle (0.07cm)
	(1,1) circle (0.07cm)
	(2,0) circle (0.07cm)
	(3,1) circle (0.07cm)
	(4,0) circle (0.07cm)
	(4,3) circle (0.07cm)
	;
\fill[blue]
	(1,2) circle (0.07cm)
	(2,3) circle (0.07cm)
	(2,1) circle (0.07cm)
	(2,2) circle (0.07cm)
	(3,2) circle (0.07cm)
	;
\draw (0,0) circle (0.07cm)
	(0,3) circle (0.07cm)
	(1,1) circle (0.07cm)
	(2,0) circle (0.07cm)
	(3,1) circle (0.07cm)
	(4,0) circle (0.07cm)
	(4,3) circle (0.07cm)
	(1,2) circle (0.07cm)
	(2,3) circle (0.07cm)
	(2,1) circle (0.07cm)
	(2,2) circle (0.07cm)
	(3,2) circle (0.07cm)
	;
\draw [orange, dashed] (2,1.5) ellipse (1.7cm and 1.1cm);
\draw [orange, dashed]  plot [smooth] coordinates {(1,-.5) (1.5,1) (1.5,2) (1,3.5)};
\draw [orange, dashed]  plot [smooth] coordinates {(3,-.5) (2.5,1) (2.5,2) (3,3.5)};
\draw [orange, dashed] (-.5,1.5)--(4.5,1.5);
\end{tikzpicture}
\captionof{figure}{\label{closed3} A well-posed IVP for 4-dimensionally consistent equations.}
\end{center}
For other quad-graphs with closed strips, see e.g. \cite[Example 6]{AV} and Figure \ref{closed3}, not only do we need to impose the same equation on all quads
(with the correct dependence on the edge parameters) the equation also needs to be multi-dimensionally consistent. For every closed strip
there is one quad not used to determine a value, but whose equation is satisfied identically given the values determined by the other quads.
Thus, for integrable equations the right number of initial values is
\[
n=v-(q-c)=s+1-h.
\]

For any quad-graph, the closed strips can be removed by removing these non-determining quads (creating holes). On the resulting graph (with no closed strips) the IVP is well-posed for generic relations on the quads. For example, by removing one edge from the boundary of the graph in Figure \ref{closed3} the IVP would be well-posed with generic relations given on the quads. It is important to realise that this only works for graphs were the closed strips are contractable, that is, no strips may close around a hole, see Figure \ref{closed4}.

\begin{center}
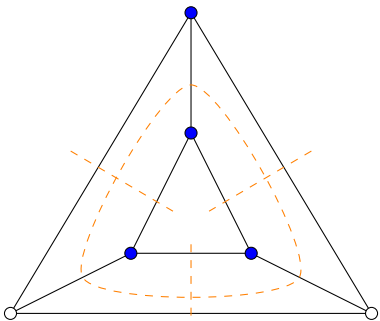

\begin{tikzpicture}[scale=.8]
\draw[thin]
    (0,0)--(3,5)--(6,0)--(0,0)--(2,1)--(4,1)--(6,0)
    (2,1)--(3,3)
    (3,5)--(3,3)--(4,1);
\fill[white] (0,0) circle (0.1cm)
	(6,0) circle (0.1cm)
	;
\fill[blue]
	(2,1) circle (0.1cm)
	(4,1) circle (0.1cm)
	(3,3) circle (0.1cm)
	(3,5) circle (0.1cm)
	;
\draw (0,0) circle (0.1cm)
	(6,0) circle (0.1cm)
	(2,1) circle (0.1cm)
	(4,1) circle (0.1cm)
	(3,3) circle (0.1cm)
	(3,5) circle (0.1cm)
	;
\draw [orange, dashed]  plot [smooth cycle] coordinates {(1.2,.6) (4.8,.6) (3,3.8) };
\draw [orange, dashed] (3,-.3)--(3,1.2)
	(1,2.7)--(2.7,1.7)
	(5,2.7)--(3.3,1.7);
\end{tikzpicture}
\parbox{11.5cm}{\captionof{figure}{\label{closed4} An over-determined IVP. Integrability does not imply consistency around a hole.}}
%By removing an edge from the boundary we do obtain a graph on which the IVP is well-posed.}}
\end{center}

An important remark, also made in \cite{AV}, is that integrable quad-equations are {\em degenerate} in the sense that
if the two lattice parameters are equal, then the equation factorises into a product of the differences over the "diagonals" of the quad. As shown by Atkinson this is in fact a property of equations which are strongly cubic consistent, see \cite{JAP}.

If the loop of the self-intersecting strip does not go around a hole, the two values on the diagonal coincide, and an integrable equation
does not determine the value of the fourth vertex from the other three on the quad where the strip self-intersects, cf.
Figure \ref {sis1}. We note that the configuration of singular edges in Figure \ref{sis1} is not admissible according to
\cite[Theorem 1]{JA}, which shows it is not only natural but also necessary to impose the mild technical
assumption, cf. \cite[Section 2.4]{JA}.

\begin{center}
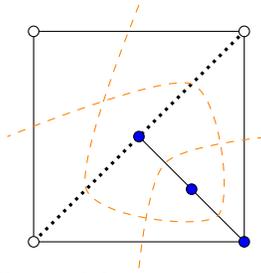

\begin{tikzpicture}[scale=.7]
\draw[thin]
    (0,0)--(0,4)--(4,4)--(4,0)--(0,0)
    (2,2)--(4,0)
   ;
\draw[dotted, very thick]
	(0,0)--(4,4);
\fill[white]
	(0,0) circle (0.1cm)
	(0,4) circle (0.1cm)
	(4,4) circle (0.1cm)
	;
\fill[blue] (4,0) circle (0.1cm)
	(2,2) circle (0.1cm)
	(3,1) circle (0.1cm)
	;
\draw
	(0,0) circle (0.1cm)
	(0,4) circle (0.1cm)
	(4,4) circle (0.1cm)
	(4,0) circle (0.1cm)
	(2,2) circle (0.1cm)
	(3,1) circle (0.1cm)
	;
\draw [orange, dashed]  plot [smooth] coordinates {(2,4.5) (1,1) (3.5,.5) (3,3) (-.5,2)};
\draw [orange, dashed]  plot [smooth] coordinates {(2,-.5) (2.5,1.5) (4.5,2)};
\end{tikzpicture}
\parbox{11.5cm}{\captionof{figure}{\label{sis1} Although well-posed for generic equations, this IVP is under-determined
for integrable equations. The solution is singular at the top-left vertex.}}
\end{center}
However, if the loop of the self-intersecting strip goes around a hole, the two values on the diagonal of the quad where the strip self-intersects do not coincide, see Figure \ref{sis2}.

\begin{center}
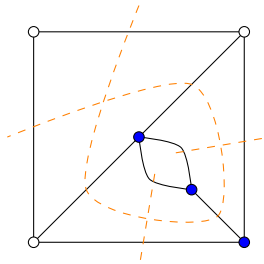

\begin{tikzpicture}[scale=.7]
\draw[thin]
    (0,0)--(0,4)--(4,4)--(4,0)--(0,0)--(4,4)
    (3,1)--(4,0)
    plot [smooth] coordinates {(3,1) (2.2,1.2) (2,2)}
    plot [smooth] coordinates {(3,1) (2.8,1.8) (2,2)}
   ;
\fill[white]
	(0,0) circle (0.1cm)
	(0,4) circle (0.1cm)
	(4,4) circle (0.1cm)
	;
\fill[blue] 	(4,0) circle (0.1cm)
	(2,2) circle (0.1cm)
	(3,1) circle (0.1cm)
	;
\draw
	(0,0) circle (0.1cm)
	(0,4) circle (0.1cm)
	(4,4) circle (0.1cm)
	(4,0) circle (0.1cm)
	(2,2) circle (0.1cm)
	(3,1) circle (0.1cm)
	;
%\draw[thick, blue]
%	(4,0)--(3,1)
%	;
%\draw [blue, thick]  plot [smooth] coordinates {(3,1) (2.2,1.2) (2,2)};
\draw [orange, dashed]  plot [smooth] coordinates {(2,4.5) (1,1) (3.5,.5) (3,3) (-.5,2)};
\draw [orange, dashed]  (2.3,1.3)--(2,-.5);
\draw [orange, dashed]  (2.7,1.7)--(4.5,2);
\end{tikzpicture}
\parbox{11.5cm}{\captionof{figure}{\label{sis2} A well-posed IVP for both integrable and non-integrable equations. Here the values at the bottom-left vertex
and the top-right vertex differ.}}
\end{center}

\section{A no-go statement}
There does not exist a local condition for the well-posedness of IVPs. What we mean is the following.
Suppose a graph $G$ has a well-posed IVP. We can glue to $G$ the graph of Figure \ref{one} so that the blue path
lies on the boundary of $G$. From the IVP on $G$ the values at its boundary can be calculated uniquely, and in particular the
values  $u,u_1,u_2$. But the values of $v,w,u_{12}$ are not uniquely determined. Therefore the IVP is not well-posed
on the extended graph. Thus any condition for well-posedness should take into account such sub-graphs which can possibly
be far away from the initial values. Figure \ref{glueon} presents two other graphs which would turn a well-posed
IVP into a nearly well-posed IVP when glued onto the boundary.

\begin{center}
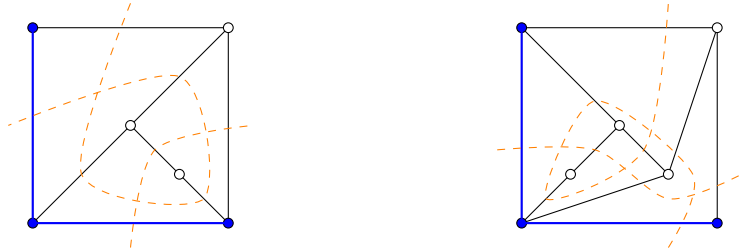

\begin{tikzpicture}[scale=.65]
\draw[thin]
    (0,0)--(0,4)--(4,4)--(4,0)--(0,0)--(4,4)
    (2,2)--(4,0)
   ;
\fill[white]
	(3,1) circle (0.1cm)
	(2,2) circle (0.1cm)
	(4,4) circle (0.1cm)
	;
\fill[blue] (0,0) circle (0.1cm)
	(0,4) circle (0.1cm)
	(4,0) circle (0.1cm)
	;
\draw
	(3,1) circle (0.1cm)
	(2,2) circle (0.1cm)
	(4,4) circle (0.1cm)
	(0,0) circle (0.1cm)
	(0,4) circle (0.1cm)
	(4,0) circle (0.1cm)
	;
\draw[thick, blue]
	(4,0)--(0,0)--(0,4)
	;
\draw [orange, dashed]  plot [smooth] coordinates {(2,4.5) (1,1) (3.5,.5) (3,3) (-.5,2)};
\draw [orange, dashed]  plot [smooth] coordinates {(2,-.5) (2.5,1.5) (4.5,2)};

\draw[thin]
	(0+10,0)--(0+10,4)--(4+10,4)--(4+10,0)--(0+10,0)--(2+10,2)--(0+10,4)
	(2+10,2)--(3+10,1)--(4+10,4)
	(3+10,1)--(0+10,0)
	;
\fill[white] (1+10,1) circle (0.1cm)
	(2+10,2) circle (0.1cm)
	(3+10,1) circle (0.1cm)
	(4+10,4) circle (0.1cm)
		;
\fill[blue]
	(0+10,0) circle (0.1cm)
	(4+10,0) circle (0.1cm)
	(0+10,4) circle (0.1cm)
	;
\draw (1+10,1) circle (0.1cm)
	(2+10,2) circle (0.1cm)
	(3+10,1) circle (0.1cm)
	(4+10,4) circle (0.1cm)
	(0+10,0) circle (0.1cm)
	(4+10,0) circle (0.1cm)
	(0+10,4) circle (0.1cm)
	;
\draw[thick, blue]
	(4+10,0)--(0+10,0)--(0+10,4)
	;
\draw [orange, dashed]  plot [smooth] coordinates {(-.5+10,1.5) (1.5+10,1.5) (3+10,.5) (4.5+10,1) };
\draw [orange, dashed] plot [smooth] coordinates {(3+10,-.5) (3.5+10,1) (1.5+10,2.5) (.5+10,.5) (2.5+10,1.5) (3+10,4.5)};

\end{tikzpicture}
\captionof{figure}{\label{glueon} Two nearly well-posed IVPs.}
\end{center}

\section{Quad-graphs with no well-posed IVP}
Not all quad-graphs admit well-posed IVPs. First of all, there are the graphs with closed strips. Although, if the strips are contractable, for integrable equation such graphs do admit well-posed IVPs. In this section we present examples without closed strips.

A quad-graph with only one strip, which does not admit a well-posed IVP is given in Figure \ref{non-simple}.

By gluing one of the graphs in Figure \ref{one} or \ref{glueon} onto itself (or onto another one) along the blue edges we obtain quad-graphs with two strips on which no well-posed IVP exists, see Figure \ref{twoc1}.

\noindent
\parbox{55mm}{
\begin{center}
\begin{tikzpicture}[scale=.75]
\draw[thin]
    (3,0)--(0,3)--(3,6)--(6,3)--(3,0)--(2,2)--(4,4)--(3,6)--(2,4)--(4,2)--(3,0) (0,3)--(6,3);
\draw[thin, blue] (0,3)--(6,3);
\fill[white] (3,0) circle (0.1cm)
	(2,2) circle (0.1cm)
	(4,2) circle (0.1cm)
	(0,3) circle (0.1cm)
	(3,3) circle (0.1cm)
	(6,3) circle (0.1cm)
	(2,4) circle (0.1cm)
	(4,4) circle (0.1cm)
	(3,6) circle (0.1cm);
\draw (3,0) circle (0.1cm)
	(2,2) circle (0.1cm)
	(4,2) circle (0.1cm)
	(0,3) circle (0.1cm)
	(3,3) circle (0.1cm)
	(6,3) circle (0.1cm)
	(2,4) circle (0.1cm)
	(4,4) circle (0.1cm)
	(3,6) circle (0.1cm);
\draw [orange, dashed] plot [smooth] coordinates {(0,2) (2,2.5) (3,2) (4,1.5) (5,3) (4,4.5) (3,4) (2,3.5) (0,4)};
\draw [orange, dashed] plot [smooth] coordinates {(6,2) (4,2.5) (3,2) (2,1.5) (1,3) (2,4.5) (3,4) (4,3.5) (6,4)};
\end{tikzpicture}
\end{center}
} \hfill \parbox{55mm}{
\begin{center}
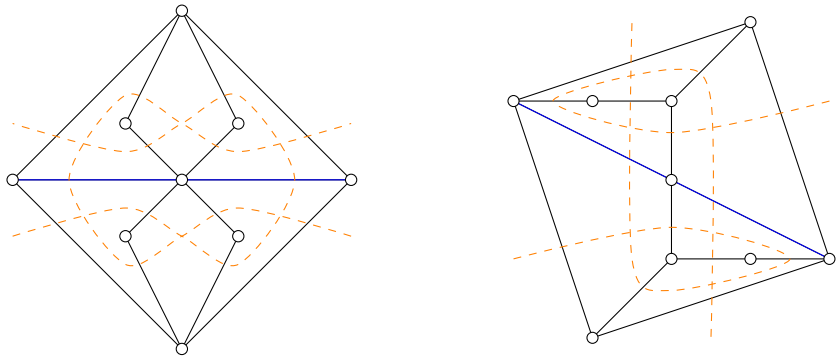

\begin{tikzpicture}[scale=1.05]
\draw[thin]
    (1,0)--(0,3)--(3,4)--(4,1)--(1,0)--(2,1)--(2,3)--(0,3)
    (4,1)--(2,1)  (2,3)--(3,4) (0,3)--(4,1);
\draw[thin, blue] (0,3)--(4,1);
\fill[white] (0,3) circle (0.07cm)
	(1,0) circle (0.07cm)
	(1,3) circle (0.07cm)
	(2,1) circle (0.07cm)
	(2,2) circle (0.07cm)
	(2,3) circle (0.07cm)
	(3,1) circle (0.07cm)	
	(3,4) circle (0.07cm)
	(4,1) circle (0.07cm)
	;
\draw (0,3) circle (0.07cm)
	(1,0) circle (0.07cm)
	(1,3) circle (0.07cm)
	(2,1) circle (0.07cm)
	(2,2) circle (0.07cm)
	(2,3) circle (0.07cm)
	(3,1) circle (0.07cm)	
	(3,4) circle (0.07cm)
	(4,1) circle (0.07cm)
	;
\draw [orange, dashed] plot [smooth] coordinates {(4,3) (2,2.6) (.5,3)(2,3.4) (2.5,3) (2.5,0)};
\draw [orange, dashed] plot [smooth] coordinates {(0,1) (2,1.4) (3.5,1)(2,.6) (1.5,1) (1.5,4)};
\end{tikzpicture}
\end{center}}
\vspace{2mm}
\captionof{figure}{\label{twoc1} Here any set of three vertices provides a nearly well-posed IVP.}

\smallskip
\noindent
An example which is not simply connected and an example with three strips are given in Figure \ref{dnawp}.

\noindent
\parbox{57mm}{
\begin{center}
\begin{tikzpicture}[scale=.5]
\draw[orange, dashed]
    (-.5,3)--(6.5,3);
\draw[thin] (3,2)--(3,6)
	(3,3) circle (3cm)
	(3,1) circle (1cm);
\fill[white] (3,0) circle (0.15cm)
	(3,2) circle (0.15cm)
	(3,4) circle (0.15cm)
	(3,6) circle (0.15cm);
\draw (3,0) circle (0.15cm)
	(3,2) circle (0.15cm)
	(3,4) circle (0.15cm)
	(3,6) circle (0.15cm);
\draw [orange, dashed] plot [smooth] coordinates {(2.3,1.3) (1.3,3) (3,5) (4.7,3)(3.7,1.3)};
\end{tikzpicture}
\end{center}
} \hfill \parbox{57mm}{
\begin{center}
\begin{tikzpicture}[scale=.9, rotate=-90]
\draw[thin]
    (3,0)--(0,4)--(1,6)--(5,4)--(3,0)
    (0,4)--(5,4)
    (1,6)--(1,4)--(3,0)--(3,5)
    (1,6)--(2,4)--(2,2)--(4,2)
    (2,3)--(3,3)--(5,4);
\fill[white] (3,0) circle (0.07cm)
	(3,1) circle (0.07cm)
	(2,2) circle (0.07cm)
	(3,2) circle (0.07cm)
	(4,2) circle (0.07cm)
	(2,3) circle (0.07cm)
	(3,3) circle (0.07cm)
	(0,4) circle (0.07cm)
	(1,4) circle (0.07cm)
	(2,4) circle (0.07cm)
	(3,4) circle (0.07cm)
	(4,4) circle (0.07cm)
	(5,4) circle (0.07cm)
	(1,5) circle (0.07cm)
	(3,5) circle (0.07cm)
	(1,6) circle (0.07cm)
	;
\draw (3,0) circle (0.07cm)
	(3,1) circle (0.07cm)
	(2,2) circle (0.07cm)
	(3,2) circle (0.07cm)
	(4,2) circle (0.07cm)
	(2,3) circle (0.07cm)
	(3,3) circle (0.07cm)
	(0,4) circle (0.07cm)
	(1,4) circle (0.07cm)
	(2,4) circle (0.07cm)
	(3,4) circle (0.07cm)
	(4,4) circle (0.07cm)
	(5,4) circle (0.07cm)
	(1,5) circle (0.07cm)
	(3,5) circle (0.07cm)
	(1,6) circle (0.07cm)
	;
\draw [orange, dashed] plot [smooth] coordinates {(-.1,3.5) (3,3.5) (4.5,4) (3,4.5) (-.2,4.5)};
\draw [orange, dashed] plot [smooth] coordinates {(2.5,5.6) (2.5,2) (3,.5) (3.5,2) (3.5,5.1)};
\draw [orange, dashed] plot [smooth] coordinates {(4.2,1.5) (2,1.7) (.6,4) (1,5.5) (2,2.7) (4.8,2.5)};

%\draw (9,1) circle (0.07cm)
%	(9,2) circle (0.07cm)
%	(9,3) circle (0.07cm)
%	(9,4) circle (0.07cm)
%	(9,5) circle (0.07cm)
%	(9,6) circle (0.07cm)
%	(7,3.5) circle (0.07cm)
%	(11,3.5) circle (0.07cm)
%	;
%\draw[thin]
%    (9,1)--(9,3)--(11,3.5)--(9,4)--(9,6)
%    (9,1)--(7,3.5)--(9,6)--(11,3.5)--(9,1)
%    (9,3)--(7,3.5)--(9,4)
%  ;
%\draw [orange, dashed] plot [smooth cycle] coordinates {(9,3.5) (8,4) (9,5.5) (10,4) (9,3.5) (8,3) (9,1.5) (10,3)};
%\draw [orange, dashed] plot [smooth] coordinates {(7.5,5) (9,4.5) (10.5,5) };
%\draw [orange, dashed] plot [smooth] coordinates {(7.5,2) (9,2.5) (10.5,2) };
\end{tikzpicture}
\end{center}
}
\hspace{2mm}
\captionof{figure}{\label{dnawp} These quad-graphs do not admit a well-posed IVP.}

\smallskip
\noindent
To describe and understand which IVPs are well-posed, or why certain IVPs
%, like the ones in Figures \ref{one} and \ref{glueon},
are nearly well-posed, we will investigate the strips and how they might entangle.

\section{Admissible configurations of intersecting curves}
We prove a correspondence between intersecting curves in perforated planes and non-simply connected proper quad-graphs.
Indeed, the curves correspond to the strips of the quad-graph, the areas enclosed by the curves correspond to the
vertices of the graph, the intersection points correspond to the quads, and the holes correspond to the holes (or connected sets of holes) of the quad-graph. A {\em path} on a set of intersecting curves is a path over curves which can turn at intersection points and go through holes. We need the following restrictions on a configuration of intersecting curves:
\begin{itemize}
\item[$\ast$] Curves intersect transversally and at most two curves intersect in one point.
\item[$\ast$] Every curve intersects and either it connects two holes, or it is closed.
\item[$\ast$] Through every hole and through every closed curve, as well as through every loop in a self-intersecting curve, there is a path that starts and ends in the outside hole and does not intersect itself.
\end{itemize}
If these conditions are satisfied the configuration is called {\em admissible}. We note that the fact that every curve intersects ensures that the holes are not connected to each other.

\noindent
\parbox{57mm}{
\begin{center}
\begin{tikzpicture}[scale=.7]
\draw (3,2) ellipse (3cm and 2cm);
\draw [orange, thick] plot [smooth] coordinates {(.6,.6) (2,2) (3.5,2.7) (5,2.1)(4.5,1)(3.5,1)(2.5,2.5)(3.5,3.5)(4.5,2.5)(2.8,1)(1.5,2.5)(2,4)};
\end{tikzpicture}
\captionof{figure}{A non-admissible configuration.}
\end{center}
} \hfill \parbox{57mm}{
\begin{center}
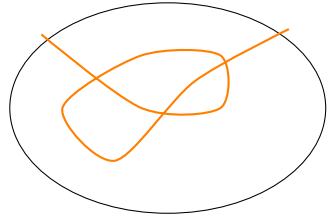

\begin{tikzpicture}[scale=.7]
\draw (3,2) ellipse (3cm and 2cm);
\draw [orange, thick] plot [smooth] coordinates {(.6,3.4) (2.5,2) (4,2) (4,3) (2.5,3) (1,2) (2,1)(3.5,2.5)(5.3,3.5) };
\end{tikzpicture}
\captionof{figure}{An admissible configuration (corresponding to the quad-graph in Figure \ref{non-simple}).}
\end{center}
}

\smallskip
\noindent
To clarify that the strips of a quad-graph form an admissible configuration of intersecting curves, we note that
non-closed strips start and end at an edge on the boundary. That every strip intersects another strip, transversally and at most two at one point, is evident as the strips are defined on quads. 

Let $C$ denote a hole, loop, or closed strip. It is clear that if $C$ is a closed strip or a loop, it consists of at least two distinct quads. If $C$ is a hole there are also at least two distinct quads adjacent to 
$C$ (if only one quad was adjacent to a hole, there would be a vertex with more than two edges bounding a single quad). Thus, we have a path through $C$. As quad-graphs are connected there is a path $P$ on the graph connecting a vertex of $C$ with a vertex on the outside boundary. As paths over strips may turn or go through holes, the path over strips through $C$ can be constructed such that it closely follows $P$ on both sides without crossing $P$, cf. Figure \ref{pathofstrips}.

\begin{center}
\begin{tikzpicture}[scale=.5]
\draw[thin]
    (1,1)--(1,4)--(3,6)--(9,6)--(11,8)--(8,8)--(6,6)
    (6,3)--(4,1)--(1,1)--(3,3)--(9,3)--(11,5)--(11,8)
    (3,3)--(3,6)
    	(6,4.5) circle (1.5cm)
   ;
\draw[thick, blue]
  (1,1)--(4,1)--(6,3)--(9,3)--(9,6)--(11,8);
\draw [orange, dashed] plot [smooth] coordinates {(1.5,0) (4.4,2.8) (4,4.5) (4.5,7)};
\draw [orange, dashed] plot [smooth] coordinates {(2,6) (2.2,2.2) (6,2)};
\draw [orange, dashed] plot [smooth] coordinates {(0.2,1.9) (3,4.5) (5.2,4.5)};
\draw [orange, dashed] plot [smooth] coordinates {(10.5,9) (7.6,6.2) (8,4.5) (7.5,2)};
\draw [orange, dashed] plot [smooth] coordinates {(10,3) (9.8,6.8) (6,7)};
\draw [orange, dashed] plot [smooth] coordinates {(12,7.3) (9,4.5) (6.8,4.5)};
\begin{scope}
\clip (0,1) rectangle (3.6, 3.6);
\draw [red, thick] plot [smooth] coordinates {(2,6) (2.2,2.2) (6,2)};
\draw [red, thick] plot [smooth] coordinates {(0.2,1.9) (3,4.5) (5.2,4.5)};
\end{scope}
\begin{scope}
\clip (3.6,1) rectangle (6, 4.7);
\draw [red, thick] plot [smooth] coordinates {(1.5,0) (4.4,2.8) (4,4.5) (4.5,7)};
\end{scope}
\begin{scope}
\clip (4,3) rectangle (6, 6);
\draw [red, thick] plot [smooth] coordinates {(0.2,1.9) (3,4.5) (5.2,4.5)};
\end{scope}
\begin{scope}
\clip (5,4.3) rectangle (11, 9);
\draw [red, thick] plot [smooth] coordinates {(10.5,9) (7.6,6.2) (8,4.5) (7.5,2)};
\end{scope}
\begin{scope}
\clip (6,3) rectangle (8, 6);
\draw [red, thick] plot [smooth] coordinates {(12,7.3) (9,4.5) (6.8,4.5)};
\end{scope}
\fill[white] (1,1) circle (0.1cm)
	(1,4) circle (0.1cm)
	(3,3) circle (0.1cm)
	(3,6) circle (0.1cm)
	(4,1) circle (0.1cm)
	(6,3) circle (0.1cm)
	(6,6) circle (0.1cm)
	(8,8) circle (0.1cm)
	(9,3) circle (0.1cm)
	(9,6) circle (0.1cm)
	(11,5) circle (0.1cm)
	(11,8) circle (0.1cm)
	;
\draw (1,1) circle (0.1cm)
	(1,4) circle (0.1cm)
	(3,3) circle (0.1cm)
	(3,6) circle (0.1cm)
	(4,1) circle (0.1cm)
	(6,3) circle (0.1cm)
	(6,6) circle (0.1cm)
	(8,8) circle (0.1cm)
	(9,3) circle (0.1cm)
	(9,6) circle (0.1cm)
	(11,5) circle (0.1cm)
	(11,8) circle (0.1cm)
	;
\end{tikzpicture}
\hspace{2mm}\parbox{11.5cm}{
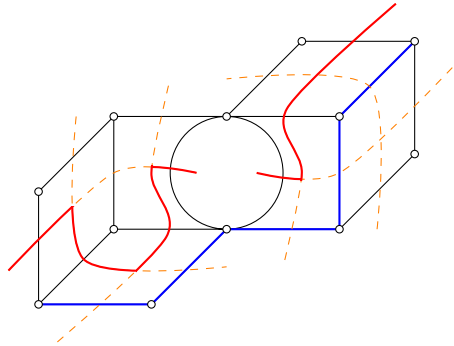
\captionof{figure}{\label{pathofstrips} A red path on a set of intersecting curves closely following a blue path on
a quad-graph.}}
\end{center}

\smallskip
\noindent
To convince the reader that any admissible configuration of curves gives rise to a quad-graph, we show that every intersection
of curves separates the plane into four different areas. At any point of intersection $p$ two curves, $C$ and $D$, intersect.
Suppose $C\neq D$. Let both $C,D$ be open, so they connect a hole to a hole. As through every hole there is a path to the outer hole that does not intersect itself, we can choose a path which does not go through $p$ again. Therefore, each of the four areas is enclosed by a path and is separated from the other areas by Jordan's curve theorem.
If $C$ or $D$ is closed a similar argument applies. If $C=D$ the intersection point, $p$, cuts the curve $C$ into two curves,
$L$ and  $R$ (which take a turn at $p$). The curves $L$ and $R$ can be open, in which case they connect two holes, or
closed, in which case a path $P$ goes through, connecting the outside hole to itself and with $p\not\in\V(P)$. In both cases, the four areas are separated, cf. Figure \ref{separation}.

\begin{center}
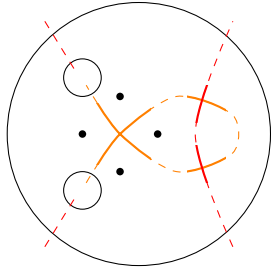

\begin{tikzpicture}[scale=.5]
\draw (3.5,4) circle (3.5cm)
	(2,5.5) circle (0.5cm)
	(2,2.5) circle (0.5cm)
	;
\fill[black] (4,4) circle (.1cm)
	(3,3) circle (.1cm)
	(3,5) circle (.1cm)
	(2,4) circle (.1cm);
\draw [orange, dashed] plot [smooth] coordinates {(2.1,2.7) (3,4) (4.5,5) (6,4.5) (6,3.5) (4.5,3) (3,4) (2.1,5.3)};
\draw [red, dashed] plot [smooth] coordinates {(6,1) (5,4) (6,7)};
\draw [red, dashed] plot [smooth] coordinates {(1,1) (1.8,2.3)};
\draw [red, dashed] plot [smooth] coordinates {(1,7) (1.8,5.7)};
\begin{scope}
\clip (2.3,3.2) rectangle (3.8, 4.8);
\draw [orange, thick] plot [smooth] coordinates {(2.1,2.7) (3,4) (4.5,5) (6,4.5) (6,3.5) (4.5,3) (3,4) (2.1,5.3)};
\draw [orange, thick] plot [smooth] coordinates {(6,1) (5,4) (6,7)};
\end{scope}
\begin{scope}
\clip (4.8,4.3) rectangle (5.8, 5.3);
\draw [orange, thick] plot [smooth] coordinates {(2.1,2.7) (3,4) (4.5,5) (6,4.5) (6,3.5) (4.5,3) (3,4) (2.1,5.3)};
\draw [red, thick] plot [smooth] coordinates {(6,1) (5,4) (6,7)};
\end{scope}
\begin{scope}
\clip (4.8,3.7) rectangle (5.8, 2.7);
\draw [orange, thick] plot [smooth] coordinates {(2.1,2.7) (3,4) (4.5,5) (6,4.5) (6,3.5) (4.5,3) (3,4) (2.1,5.3)};
\draw [red, thick] plot [smooth] coordinates {(6,1) (5,4) (6,7)};
\end{scope}
\end{tikzpicture}
\hspace{2mm}\parbox{11.5cm}{\captionof{figure}{\label{separation} A separating configuration of curves. Here $L$ (on the left) is open, and $R$ (on the right) is a closed loop, through which a red path connects the outer hole to itself.}}
\end{center}

\section{Constructing well-posed IVPs}
As we argued in section 3, in the case of general relations on the quads there are no well-posed IVPs on quad-graphs with closed strips. Moreover, the strategy we adopt for integrable equations on quad-graphs with contractable (and non-self intersecting) closed strips is to remove the closed strips by perforating the graph, i.e. for every closed strip we replace one quad by a hole. Therefore, in the sequel we assume that $c=0$.

The idea of setting up a well-posed IVP for a graph $G$ is to add vertices to the IVP one by one.
According to section \ref{rinu} we need to add $s+1-h$ values. We will choose to add the vertices so that after having added the $i$-th value, there is a connected sub-graph $G_i$ of $G$ on which the sub-IVP which assigns the first $i$ values is well-posed. We provide three obstructions for an IVP to be well-posed and
it will become clear that if these obstructions are avoided the IVP is well-posed.

Instead of adding vertices we will be colouring curves.   
%We can mimic solving an IVP on intersecting curves. This we do by colour-coding the curves.
By a {\em piece of a curve} we mean the set of all points on a curve that are in between two consecutive intersection points (or in between the intersection point closest to a hole and that hole). Thus pieces of curves are dual to edges of quad-graphs. We propose the following encoding of initial, or known, values.
\begin{itemize}
\item[Rule 0:] Colour a piece of curve green if it crosses an edge whose vertices are known.
\end{itemize}
If three vertices are known on a quad then there are two green pieces of curve which meet each other in the intersection point of the curves on the quad. Using the equation to find the value on the fourth vertex of the quad is dually represented by the
following rule.
\begin{itemize}
\item[Rule 1:] If two green pieces of intersecting curves meet in a point then the green colour propagates through.
\end{itemize}
Given a simply connected quad-graph with a well-posed IVP on a Cauchy sub-graph we can encode
the position of its initial date by colouring one piece of each curve, using Rule 0. It should be clear to the reader that Rule 1 can then be used to colour all curves green.

For non-simply connected quad-graphs a second rule is evident.
\begin{itemize}
\item[Rule 2:] If all curves, but one, entering (or exiting) a hole are green the remaining one should be coloured green as well.
\end{itemize}
It might be instructive to colour the orange curves in Figure \ref{hole} green (or any other colour available), using Rules 0,1 and 2. 

Note that an IVP is over-determined if green propagates (either over one or over two pieces) into an intersection point where two green pieces of curve meet (cf. Figure \ref{closed4}, and Figure \ref{seven} with three initial values given). Also, if green propagates over a curve towards a quad from two sides, i.e. the
left two vertices are determined from the left and the right vertices will be determined from the right, then this
yields a contradicting on the quad, the IVP is overdetermined.

On the other hand, if green propagates over two pieces into one and the same hole, and all other pieces
entering the hole are green already, one can not apply Rule 2 at this hole but there is no inconsistency, as
there are no relations on the holes. Usually this happens at the outside hole, e.g. consider the configuration
of two curves with one intersection. However, it may also happen at an inside hole, see Figure \ref{IH}.

\begin{center}
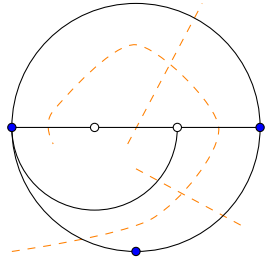

\begin{tikzpicture}[scale=1.1]
\draw[orange, dashed]
    (1.5,1)--(2.8,.3)
    (1.4,1.3)--(2.3,3);
\draw [orange, dashed] plot [smooth] coordinates {(0,0) (1.7,.4) (2.5,1.5) (1.5,2.5) (.5,1.7) (.5,1.3)};
\begin{scope}
\clip (0,0) rectangle (2, 1.5);
\draw (1,1.5) circle (1cm);
\end{scope}
\draw
	(1.5,1.5) circle (1.5cm)
	(0,1.5)--(3,1.5);
\fill[white] (1,1.5) circle (0.05cm)
	(2,1.5) circle (0.05cm);
\fill[blue] (0,1.5) circle (0.05cm)
	(3,1.5) circle (0.05cm)
	(1.5,0) circle (0.05cm)
	;
\draw (1,1.5) circle (0.05cm)
	(2,1.5) circle (0.05cm)
	(0,1.5) circle (0.05cm)
	(3,1.5) circle (0.05cm)
	(1.5,0) circle (0.05cm)
	;
\end{tikzpicture}
\hspace{2mm}\parbox{11.5cm}{\captionof{figure}{\label{IH} A well-posed IVP on a graph with one hole of degree three. We apply Rule 2 at the outside hole, instead of at the inside hole.}}
\end{center}

The strategy for posing the IVP will be as follows. First step is to choose an enclosed area (this corresponds to
choosing the first vertex, which is denoted $G_1$). Secondly we colour one piece of the enclosure green. The maximal sub-graph $G_2$ consists of two vertices and one edge, or in the dual picture: two areas separated by a green piece of curve and enclosed by orange pieces of curve. Obviously we cannot apply Rule 1 yet. So we go to step 3 and colour a piece of the enclosure of $G_2$ green. Now we might be able to propagate the green colour. This we do until we can't go any further. At every next step we colour one piece of the enclosure green and then propagate the green using Rules 0,1, or 2 exhaustively.

By colouring, at step $i$, a piece of the enclosure of $G_{i-1}$ we ensure that $G_i$ is connected.
It is important to realise that some pieces of the enclosure may belong to curves partly coloured green,
while other orange pieces belong to entirely orange curves. The difference is crucial and therefore we
propose to change the colour of the orange pieces of a partially green curve red. If we would colour a
red enclosing piece green, and if at the end the whole curve is green (all vertices of the strip are determined) then there is a quad on the strip whose vertices are determined either from the left or
from the right yielding a contradiction at this quad. If at the end the curve is not entirely green, the IVP is also not well-posed. Thus we arrive at the following rule.
\begin{itemize}
\item[Rule 3:] 
Build up the IVP by colouring green orange enclosing pieces, no red ones. 
\end{itemize}

We are almost finished answering the first question; which vertices to choose. The next rule is not necessary (and neither it is sufficient) but it doesn't hurt and in certain cases (examples will be given later) it provides the right choice. 
\begin{itemize}
\item[Rule 4:] 
Build up the IVP by colouring green orange enclosing pieces which are connected to green pieces.
\end{itemize}
Obviously this rule can't be applied in step 1 or 2 but otherwise it ensures that Rule 1 can be applied
and so $G_i$ will be larger, for each $i$.

Before turning to the obstructions for well-posedness we remark that since setting up the IVP is a $(s+1-h)$-step process, and we do not colour a piece at step 1, we apply Rule 0 $s-h$ times to colour at most $s-h$ curves. Obviously, the remaining $h$ curves should be coloured green as well. This is not necessarily done by applying Rule 2, sometimes Rule 0 has to be used as shown in Figure \ref{rule0}. Note that Rule 2 is actually a special case of Rule 0.

\begin{center}
\vspace{1mm}
\begin{tikzpicture}[scale=1.2, rotate=-90]
\draw (0,0)--(0,4)--(3,4)--(3,0)--(0,0)--(1,2)--(1,3)--(0,4)
	(1,3)--(3,2)--(1,2);
%\draw[very thick] (1,3)--(3,2);	
\begin{scope}
\clip (1,0) rectangle (3, 2);
\draw (2,2) circle (1cm);
\end{scope}
\fill[white] (2,2) circle (0.05cm)
	(3,2) circle (0.05cm)
	(1,3) circle (0.05cm)
	(3,4) circle (0.05cm)
	(0,0) circle (0.05cm)
	(1,2) circle (0.05cm)
	(3,0) circle (0.05cm)
	(0,4) circle (0.05cm);
\fill[blue] (0,0) circle (0.05cm)
	(1,2) circle (0.05cm)
	(3,0) circle (0.05cm)
	(0,4) circle (0.05cm)
	;
\draw (2,2) circle (0.05cm)
	(3,2) circle (0.05cm)
	(1,3) circle (0.05cm)
	(3,4) circle (0.05cm)
	(0,0) circle (0.05cm)
	(1,2) circle (0.05cm)
	(3,0) circle (0.05cm)
	(0,4) circle (0.05cm)
	(0,0) circle (0.05cm)
	(1,2) circle (0.05cm)
	(3,0) circle (0.05cm)
	(0,4) circle (0.05cm)
	;
\draw [orange, dashed] plot [smooth] coordinates {(2,4.3) (1.5,2) (1.6,1.7)};
\draw [red, dashed] plot [smooth] coordinates {(-.3,3) (2.5,2) (2.5,1.7) };
\draw [red, dashed] plot [smooth] coordinates {(3.2,3) (.6,3.4) (.5,1) (3.2,.6)};
\draw [green, thick] (1.7,-.2)--(2,1.3);
\begin{scope}
\clip (-.3,0) rectangle (1.57, 4);
\draw [green, thick] plot [smooth] coordinates {(-.3,3) (2.5,2) (2.5,1.7) };
\end{scope}
\begin{scope}
\clip (0,0) rectangle (1.78, 4);
\draw [green, thick] plot [smooth] coordinates {(3.2,3) (.6,3.4) (.5,1) (3.2,.6)};
\end{scope}
\begin{scope}
\clip (0,0) rectangle (4, 2);
\draw [green, thick] plot [smooth] coordinates {(3.2,3) (.6,3.4) (.5,1) (3.2,.6)};
\end{scope}
%\node at (1.2,3.2) {$A$};
%\node at (3.2,2.2) {$B$};
\end{tikzpicture}
\vspace{-9mm}
\hspace{2mm}\parbox{11.5cm}{
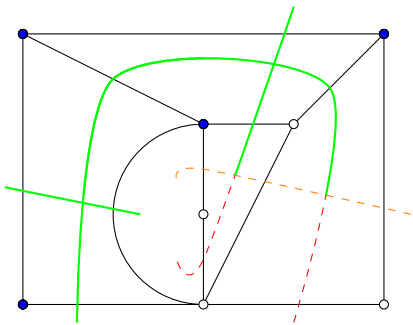
\captionof{figure}{\label{rule0} Rule 0 should be used to colour the middle piece of the orange curve, as both vertices
of the edge it crosses are known. In exchange Rule 2 is not applied at the inside or outside hole. The IVP is well-posed.}}
\end{center}

Below we provide three obstructions to well-posedness.
\begin{itemize}
\item[Obstruction 1:] 
If Rule 0 or 2 cannot be applied $h$ times.
\end{itemize}
This is the case if the connected sub-graph $G_{s+1-h}$ does not enclose an inside hole.
An example is given in Figure \ref{O1}.

\begin{center}
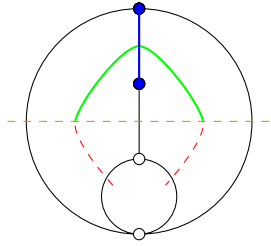

\begin{tikzpicture}[scale=.5]
\draw[thin] (3,2)--(3,4)
	(3,3) circle (3cm)
	(3,1) circle (1cm)
	(3,6)--(3,4);
\fill[white] (3,0) circle (0.15cm)
	(3,2) circle (0.15cm)
	(3,4) circle (0.15cm)
	(3,6) circle (0.15cm);
\fill[blue] (3,6) circle (0.15cm)
	(3,4) circle (0.15cm);
\draw (3,0) circle (0.15cm)
	(3,2) circle (0.15cm)
	(3,4) circle (0.15cm)
	(3,6) circle (0.15cm)
	(3,6) circle (0.15cm)
	(3,4) circle (0.15cm);
\draw[orange, dashed]
    (-.5,3)--(6.5,3);
\draw [red, dashed] plot [smooth] coordinates {(2.3,1.3) (1.3,3) (3,5) (4.7,3)(3.7,1.3)};
\begin{scope}
\clip (0,3) rectangle (6, 6);
\draw [green, thick] plot [smooth] coordinates {(2.3,1.3) (1.3,3) (3,5) (4.7,3)(3.7,1.3)};
\end{scope}
\draw[blue, thick] (3,4)--(3,6);
\end{tikzpicture}
\hspace{2mm}\parbox{11.5cm}{\captionof{figure}{\label{O1} The sub-graph $G_2$ (blue) does not enclose the hole. The green does not propagate.}}
\end{center}

\begin{itemize}
\item[Obstruction 2:] 
If by applying Rule 0 or 2 a red piece is coloured green. An example is given in Figure \ref{O2}.
\end{itemize}

\begin{center}
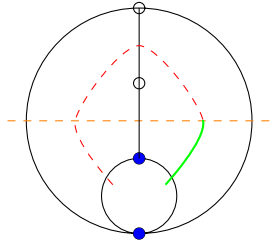

\begin{tikzpicture}[scale=.5]
\draw (3,0) circle (0.15cm)
	(3,2) circle (0.15cm)
	(3,4) circle (0.15cm)
	(3,6) circle (0.15cm)
	(3,3) circle (3cm)
	(3,1) circle (1cm);
\fill[blue] (3,2) circle (0.15cm)
	(3,0) circle (0.15cm);
\draw[orange, dashed]
    (-.5,3)--(6.5,3);
\draw[thin] (3,2)--(3,6);
\begin{scope}
\clip (0,0) rectangle (3, 3);
\draw [red, dashed] plot [smooth] coordinates {(2.3,1.3) (1.3,3) (3,5) (4.7,3)(3.7,1.3)};
\end{scope}
\begin{scope}
\clip (0,3) rectangle (6, 6);
\draw [red, dashed] plot [smooth] coordinates {(2.3,1.3) (1.3,3) (3,5) (4.7,3)(3.7,1.3)};
\end{scope}
\begin{scope}
\clip (3,0) rectangle (6, 3);
\draw [green, thick] plot [smooth] coordinates {(2.3,1.3) (1.3,3) (3,5) (4.7,3)(3.7,1.3)};
\end{scope}
\end{tikzpicture}
\hspace{2mm}\parbox{11.5cm}{\captionof{figure}{\label{O2} We now use Rule 2 to colour a red piece green. This implies one curve stays orange. The green does not propagate.}}
\end{center}

\begin{itemize}
\item[Obstruction 3:] 
If, for some $i$, the curves leaving $G_i$ entangle.
\end{itemize}

Suppose there is a curve $C_0$ leaving $G_i$ through a pair of red enclosing pieces $P_1$, where $C_0$ changes colour from green to red. The only way for the curve $C_0$ to become green entirely is that part of $P_1$ gets coloured green enabling the green from $C_0$ in $G_i$ to propagate through $P_1$. So we consider the curve $C_1$ that contains $P_1$. Starting from $P_1$ we track $C_1$ back to where it crosses the boundary of $G_i$. If $C_1$ leaves the boundary of $G_i$ through orange pieces, these pieces can be coloured green at a later step and then green may propagate through to $P_1$. If $C_2$ leaves $G_i$ through a pair of red pieces, $P_2$, we repeat the argument as for $P_1$; define $C_2$ as the curve that contains $P_2$ et cetera. If we keep on entering $G_i$ through pairs of red pieces, and for some $j$ we have $P_{j+1}=P_1$ (and hence $C_j=C_0$), then none of the red pieces of the curves $C$ will be coloured green in the process. We call the set of curves $\{C_1,C_2,\ldots,C_j\}$ {\em entangled}. We have now shown that if for some $i$ the set of curves
leaving $G_i$ contains an entangled set, then the IVP is nearly well-posed.

\bigskip
We illustrate the method by setting up an IVP on the quad-graph in Figure \ref{two}.
In step 1 we choose the top vertex as our first initial value, the enclosing boundary is drawn with fat orange lines in Figure \ref{s1}. There are two options available now, we can colour green either one of the two orange pieces of boundary.

\noindent
\parbox{40mm}{
\begin{center}
\begin{tikzpicture}[scale=1.2, rotate=-90]
\draw[thin]
    (0,1)--(1.5,2)--(1,1)--(1.5,0)--(2,1)--(1.5,2)--(3,1)--(1.5,0)--(0,1);
\fill[white] (1,1) circle (0.05cm)
	(2,1) circle (0.05cm)
	(1.5,2) circle (0.05cm)
	(1.5,0) circle (0.05cm)
	(3,1) circle (0.05cm);
\fill[blue]
	(0,1) circle (0.05cm);
\draw (1,1) circle (0.05cm)
	(2,1) circle (0.05cm)
	(1.5,2) circle (0.05cm)
	(1.5,0) circle (0.05cm)
	(3,1) circle (0.05cm)
	(0,1) circle (0.05cm);
\draw [orange, dashed] plot [smooth] coordinates {(0,0) (.5,1) (1.25,1.5) (1.5,1) (1.75,.5) (2.5,1) (3,2)};
\draw [orange, dashed] plot [smooth] coordinates {(0,2) (.5,1) (1.25,.5) (1.5,1) (1.75,1.5) (2.5,1) (3,0)};
\begin{scope}
\clip (0,0) rectangle (.5, 2);
\draw [orange, thick] plot [smooth] coordinates {(0,0) (.5,1) (1.25,1.5) (1.5,1) (1.75,.5) (2.5,1) (3,2)};
\draw [orange, thick] plot [smooth] coordinates {(0,2) (.5,1) (1.25,.5) (1.5,1) (1.75,1.5) (2.5,1) (3,0)};
\end{scope}
\end{tikzpicture}
\captionof{figure}{\label{s1} Step 1.}
\end{center}
} \parbox{40mm}{
\begin{center}
\begin{tikzpicture}[scale=1.2,rotate=-90]
\draw[thin]
    (0,1)--(1.5,2)--(1,1)--(1.5,0)--(2,1)--(1.5,2)--(3,1)--(1.5,0)--(0,1);
\fill[white] (1,1) circle (0.05cm)
	(2,1) circle (0.05cm)
	(1.5,2) circle (0.05cm)
	(3,1) circle (0.05cm);
\fill[blue]
	(1.5,0) circle (0.05cm)
	(0,1) circle (0.05cm);
\draw (1,1) circle (0.05cm)
	(2,1) circle (0.05cm)
	(1.5,2) circle (0.05cm)
	(3,1) circle (0.05cm)
	(1.5,0) circle (0.05cm)
	(0,1) circle (0.05cm);
\begin{scope}
\clip (0,0) rectangle (.5, 2);
\draw [green, thick] plot [smooth] coordinates {(0,0) (.5,1) (1.25,1.5) (1.5,1) (1.75,.5) (2.5,1) (3,2)};
\draw [orange, thick] plot [smooth] coordinates {(0,2) (.5,1) (1.25,.5) (1.5,1) (1.75,1.5) (2.5,1) (3,0)};
\end{scope}
\begin{scope}
\clip (.5, 1) rectangle (3,0);
\draw [orange, thick] plot [smooth] coordinates {(0,2) (.5,1) (1.25,.5) (1.5,1) (1.75,1.5) (2.5,1) (3,0)};
\draw [red, thick] plot [smooth] coordinates {(0,0) (.5,1) (1.25,1.5) (1.5,1) (1.75,.5) (2.5,1) (3,2)};
\end{scope}
\begin{scope}
\clip (.5, 1) rectangle (3,2);
\draw [orange,dashed] plot [smooth] coordinates {(0,2) (.5,1) (1.25,.5) (1.5,1) (1.75,1.5) (2.5,1) (3,0)};
\draw [red, dashed] plot [smooth] coordinates {(0,0) (.5,1) (1.25,1.5) (1.5,1) (1.75,.5) (2.5,1) (3,2)};
\end{scope}
\end{tikzpicture}
\captionof{figure}{\label{s2} Step 2.}
\end{center}
}\parbox{40mm}{
\begin{center}
\begin{tikzpicture}[scale=1.2, rotate=-90]
\draw[thin]
    (0,1)--(1.5,2)--(1,1)--(1.5,0)--(2,1)--(1.5,2)--(3,1)--(1.5,0)--(0,1);
\fill[white] (1,1) circle (0.05cm)
	(2,1) circle (0.05cm)
	(1.5,2) circle (0.05cm)
	;
\fill[blue]
	(1.5,0) circle (0.05cm)
	(0,1) circle (0.05cm)
	(3,1) circle (0.05cm);
\draw (1,1) circle (0.05cm)
	(2,1) circle (0.05cm)
	(1.5,2) circle (0.05cm)
	(1.5,0) circle (0.05cm)
	(0,1) circle (0.05cm)
	(3,1) circle (0.05cm);
\begin{scope}
\clip (0,0) rectangle (.5, 2);
\draw [green, thick] plot [smooth] coordinates {(0,0) (.5,1) (1.25,1.5) (1.5,1) (1.75,.5) (2.5,1) (3,2)};
\draw [red, thick] plot [smooth] coordinates {(0,2) (.5,1) (1.25,.5) (1.5,1) (1.75,1.5) (2.5,1) (3,0)};
\end{scope}
\begin{scope}
\clip (.5, 1) rectangle (2.5,0);
\draw [red, thick] plot [smooth] coordinates {(0,2) (.5,1) (1.25,.5) (1.5,1) (1.75,1.5) (2.5,1) (3,0)};
\draw [red, thick] plot [smooth] coordinates {(0,0) (.5,1) (1.25,1.5) (1.5,1) (1.75,.5) (2.5,1) (3,2)};
\end{scope}
\begin{scope}
\clip (.5, 1) rectangle (2.5,2);
\draw [red, dashed] plot [smooth] coordinates {(0,2) (.5,1) (1.25,.5) (1.5,1) (1.75,1.5) (2.5,1) (3,0)};
\draw [red, dashed] plot [smooth] coordinates {(0,0) (.5,1) (1.25,1.5) (1.5,1) (1.75,.5) (2.5,1) (3,2)};
\end{scope}
\begin{scope}
\clip (2.5,0) rectangle (3, 2);
\draw [red, thick] plot [smooth] coordinates {(0,0) (.5,1) (1.25,1.5) (1.5,1) (1.75,.5) (2.5,1) (3,2)};
\draw [green, thick] plot [smooth] coordinates {(0,2) (.5,1) (1.25,.5) (1.5,1) (1.75,1.5) (2.5,1) (3,0)};
\end{scope}
\end{tikzpicture}
\captionof{figure}{\label{s3} Step 3.}
\end{center}
}

We colour the left piece of the boundary green. This corresponds to adding the left vertex to the set of initial values, see
Figure \ref{s2}, where the boundary now contains a red piece. At the third step we can choose to colour green any of the three
orange boundary piece. If we would choose the upper or the middle one, then by repeated application of rule 1 both curves will turn
green and the corresponding IVPs will be well-posed. However choosing to colour the bottom boundary piece green we get
an entangled set of curves, as in Figure \ref{s3}. Rule 4 rules out this option.

\begin{theorem}
If an IVP is posed by colouring green one orange enclosing piece of $G_i$, for $i=1,2,\ldots s-h$, and Rule 0 (or 2) is applied $h$ times to colour an orange curve green, and no set of curves leaving $G_i$ entangles, for all $i$, then the IVP is well-posed.
\end{theorem}

{\bf Proof:} Clearly, at the end all curves are green. \hfill $\square$

\bigskip
\noindent
On simply connected quad-graphs the art of finding a well-posed IVP is to avoid entanglement of strips. A class of quad-graphs where strips do not entangle is the class of rhombic embeddable quad-graphs. A {\em rhombic embedding} of a (simply-connected) quad graph is an embedding in the plane with the property that all edges are line segments and have length 1 (and hence each bounded face is a rhombus). We have the following result \cite[Theorem 3.1]{KS}\footnote{strips were called train-tracks in \cite{KS}}
\begin{quote}
A simply-connected quad-graph has a rhombic embedding in the plane if and only if the following three
conditions are satisfied.
\begin{enumerate}
\item No strip crosses itself.
\item No strip is closed.
\item Two distinct strips cross each other at most once.
\end{enumerate}
\end{quote}
It follows from the result in section 6 that the first two conditions are a consequence of the last one (in the setting of simply-connected quad-graphs, which is the setting adopted in \cite{KS}). For rhombic embeddable  quad-graphs strips do not entangle as for any entangled set of curves $\{C_1,C_2,\ldots,C_j\}$ any pair of consecutive curves $C_i,C_{i+1}$ will intersect each other twice. This is illustrated in Figure \ref{cs}, where $C_{i+1}$ enters, by intersecting $C_i$ through $P_{i+1}$, an area defined by $C_i$ and the enclosure of the green part of the graph. Because the graph is simply connected $C_{i+1}$ needs to go into the outer hole. It does not leave the area through the enclosure (otherwise it would become part of the enclosure) so it intersects $C_i$ a second time.

\begin{center}
\begin{tikzpicture}[scale=.9]
\fill [green!20] (2.9,3.6) ellipse (2.1cm and 1.5cm);
\draw [red, thick, dashed] plot [smooth] coordinates {(1.5,5.5) (2.5,5.7)(4,5.5)(5.3,5.2)(6.7,5.6)(8,4.7)(8,3.2)(7.2,2.4)(5.5,3)(4,3)};
\draw [red, thick, dashed] plot [smooth] coordinates {(7.5,6.6)(6.7,5.6)(5.5,4)(5.5,3)(5,2)(4.2,1.5)(3.5,.7)(2.2,.5)(1.2,.9)(1.3,2)(2,3)};
\draw [red, thick] plot [smooth] coordinates {(.4,2.8)(1.3,2)(2.5,1.6)};
\draw [red, thick, dashed] plot [smooth] coordinates {(2.5,6.5)(2.5,5.7)(2.3,4.8)};
\begin{scope}
\clip (1.3,2) rectangle (2.5,5.7);
\draw [green, thick] plot [smooth] coordinates {(2.5,6.5)(2.5,5.7)(2.3,4.8)};
\draw [green, thick] plot [smooth] coordinates {(7.5,6.6)(6.7,5.6)(5.5,4)(5.5,3)(5,2)(4.2,1.5)(3.5,.7)(2.2,.5)(1.2,.9)(1.3,2)(2,3)};
\end{scope}
\begin{scope}
\clip (1.5,5) rectangle (4,6);
\draw [red, thick] plot [smooth] coordinates {(1.5,5.5) (2.5,5.7)(4,5.5)(5.3,5.2)(6.7,5.6)(8,4.7)(8,3.2)(7.2,2.4)(5.5,3)(4,3)};
\end{scope}
\begin{scope}
\clip (3,2.5) rectangle (5.5,3.5);
\draw [green, thick] plot [smooth] coordinates {(1.5,5.5) (2.5,5.7)(4,5.5)(5.3,5.2)(6.7,5.6)(8,4.7)(8,3.2)(7.2,2.4)(5.5,3)(4,3)};
\end{scope}
\begin{scope}
\clip (4.5,2) rectangle (6,4);
\draw [red, thick] plot [smooth] coordinates {(7.5,6.6)(6.7,5.6)(5.5,4)(5.5,3)(5,2)(4.2,1.5)(3.5,.7)(2.2,.5)(1.2,.9)(1.3,2)(2,3)};
\end{scope}
\draw	(1.7,5.9) node {$P_i$};
\draw	(3,6.5) node {$C_{i-1}$};
\draw	(.5,2.1) node {$P_{i+2}$};
\draw	(5.7,2.1) node {$P_{i+1}$};
\draw	(4.2,.6) node {$C_{i+1}$};
\draw	(7.7,3.5) node {$C_{i}$};
\end{tikzpicture}
\hspace{2mm}\parbox{11.5cm}{
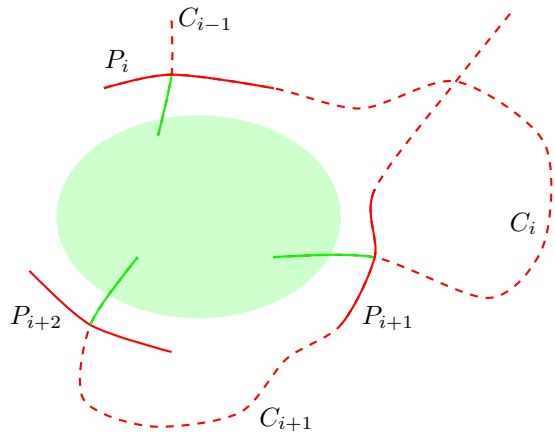
\captionof{figure}{\label{cs} Part of an entangled set of curves, showing two consecutive curves intersect each other twice.}}
\end{center}
\section{A well-posed IVP for integrable equations on the deltoidal trihexagonal tiling.}
Consider the middle rectangular band in Figure \ref{dtt}. From the seven blue points in the interior of the parallelogram
in the middle (which comprises four triangles) all points on its boundary can be calculated. By adding the point in the
centre of a triangle on the left or right,  all points of the boundary of the triangle are uniquely determined. Continuing
adding centres of triangles all values on the vertices on the rectangular band (including the ones on its boundary) can be found. Note
that none of the strips on the band are closed, so up till now we did not use the integrability of the equation. Now we 
add the centres of two triangles (each comprising four kites) which touch each other in one point and both have an edge
in common with the band. Due to multi-dimensional consistency all points on a trapezium (formed by the two
triangles and a third one) are uniquely determined. We can add centres of triangles to the left and right and widen the trapezium
till the whole band, above or under the middle one, is exhausted. Thus bands can be added on top and below to find
a solution on the full deltoidal trihexagonal tiling.

\begin{center}
\begin{tikzpicture}[scale=1.5]
\foreach \y in {0, 1, 2, 3, 4, 5}
\draw (0,\y*1.73)--(7,\y*1.73);

\draw
	(0,4*1.73)--(1,5*1.73)
	(0,2*1.73)--(3,5*1.73)
	(0,0)--(5,5*1.73)
	(2,0)--(7,5*1.73)
	(4,0)--(7,3*1.73)
	(6,0)--(7,1.73)
	(2,0)--(0,2*1.73)
	(4,0)--(0,4*1.73)
	(6,0)--(1,5*1.73)
	(7,1.73)--(3,5*1.73)
	(7,3*1.73)--(5,5*1.73);
	
\foreach \x in {1, 3, 5, 7}
\draw (\x,0)--(\x,1/1.73)
	(\x,2*1.73-1/1.73)--(\x,2*1.73+1/1.73)
	(\x,4*1.73-1/1.73)--(\x,4*1.73+1/1.73);
\foreach \x in {0, 2, 4, 6}
\foreach \y in {1, 3}
	\draw (\x,\y*1.73-1/1.73)--(\x,\y*1.73+1/1.73);
\foreach \x in {0, 2, 4, 6}
	\draw (\x,5*1.73-1/1.73)--(\x,5*1.73);
\foreach \y in {0,2,4}
\draw	
		(0,\y*1.73+1.73-1/1.73)--(.5,\y*1.73+1.73/2)--(1,\y*1.73+1/1.73)--(1.5,\y*1.73+1.73/2)--(2,\y*1.73+1.73-1/1.73)--(2.5,\y*1.73+1.73/2)--(3,\y*1.73+1/1.73)--(3.5,\y*1.73+1.73/2)--(4,\y*1.73+1.73-1/1.73)--(4.5,\y*1.73+1.73/2)--(5,\y*1.73+1/1.73)--(5.5,\y*1.73+1.73/2)--(6,\y*1.73+1.73-1/1.73)--(6.5,\y*1.73+1.73/2)--(7,\y*1.73+1/1.73);
\foreach \y in {1,3}
\draw	
		(0,\y*1.73+1/1.73)--(.5,\y*1.73+1.73/2)--(1,\y*1.73+1.73-1/1.73)--(1.5,\y*1.73+1.73/2)--(2,\y*1.73+1/1.73)--(2.5,\y*1.73+1.73/2)--(3,\y*1.73+1.73-1/1.73)--(3.5,\y*1.73+1.73/2)--(4,\y*1.73+1/1.73)--(4.5,\y*1.73+1.73/2)--(5,\y*1.73+1.73-1/1.73)--(5.5,\y*1.73+1.73/2)--(6,\y*1.73+1/1.73)--(6.5,\y*1.73+1.73/2)--(7,\y*1.73+1.73-1/1.73);

\foreach \x in {0, 1, 2, 3, 4, 5, 6, 7}
\foreach \y in {0, 1, 2, 3, 4, 5}
\fill[white] (\x, \y*1.73) circle (0.05cm);

\foreach \x in {1, 3, 5, 7}
\foreach \y in {0,  2, 4}
\fill[white] (\x, \y*1.73+1/1.73) circle (0.05cm);

\foreach \x in {1, 3, 5, 7}
\foreach \y in {1, 3}
\fill[white] (\x, \y*1.73+1.73-1/1.73) circle (0.05cm);

\foreach \x in {0, 1, 2, 3, 4, 5, 6}
\foreach \y in {0, 1, 2, 3, 4}
\fill[white] (\x+.5, \y*1.73+1.73/2) circle (0.05cm);

\foreach \x in {0, 2, 4, 6}
\foreach \y in {1,  3}
\fill[white] (\x, \y*1.73+1/1.73) circle (0.05cm);

\foreach \x in {0, 2, 4, 6}
\foreach \y in {0, 2, 4}
\fill[white] (\x, \y*1.73+1.73-1/1.73) circle (0.05cm);

\foreach \x in {1, 3, 5, 7}
{
\fill[blue] (\x, 2*1.73+1/1.73) circle (0.05cm);
\fill[blue] (\x, 2*1.73-1/1.73) circle (0.05cm);
\fill[blue] (\x, 4*1.73+1/1.73) circle (0.05cm);
}

\foreach \x in {0, 2, 4, 6}
{
\fill[blue] (\x, 3*1.73+1/1.73) circle (0.05cm);
\fill[blue] (\x, 3*1.73-1/1.73) circle (0.05cm);
\fill[blue] (\x, 1.73-1/1.73) circle (0.05cm);
}

\foreach \x in {2,3,4}
\fill[blue] (\x+.5, 2.5*1.73) circle (0.05cm);

\foreach \x in {0, 1, 2, 3, 4, 5, 6, 7}
\foreach \y in {0, 1, 2, 3, 4, 5}
\draw (\x, \y*1.73) circle (0.05cm);

\foreach \x in {1, 3, 5, 7}
\foreach \y in {0,  2, 4}
\draw (\x, \y*1.73+1/1.73) circle (0.05cm);

\foreach \x in {1, 3, 5, 7}
\foreach \y in {1, 3}
\draw (\x, \y*1.73+1.73-1/1.73) circle (0.05cm);

\foreach \x in {0, 1, 2, 3, 4, 5, 6}
\foreach \y in {0, 1, 2, 3, 4}
\draw (\x+.5, \y*1.73+1.73/2) circle (0.05cm);

\foreach \x in {0, 2, 4, 6}
\foreach \y in {1,  3}
\draw (\x, \y*1.73+1/1.73) circle (0.05cm);

\foreach \x in {0, 2, 4, 6}
\foreach \y in {0, 2, 4}
\draw (\x, \y*1.73+1.73-1/1.73) circle (0.05cm);

\begin{scope}
\clip (-.1,-.1) rectangle (7.1, 5*1.73+.1);
\foreach \x in {-2, 0, 2, 4, 6,8}
\foreach \y in {0,2,4}
\draw[orange, dashed] (\x,\y*1.73) circle (1.45cm);
\foreach \x in {-1,1,3,5,7,9}
\foreach \y in {1,3,5}
\draw[orange, dashed] (\x,\y*1.73) circle (1.45cm);
\end{scope}

\end{tikzpicture}
\hspace{2mm}\parbox{11.5cm}{
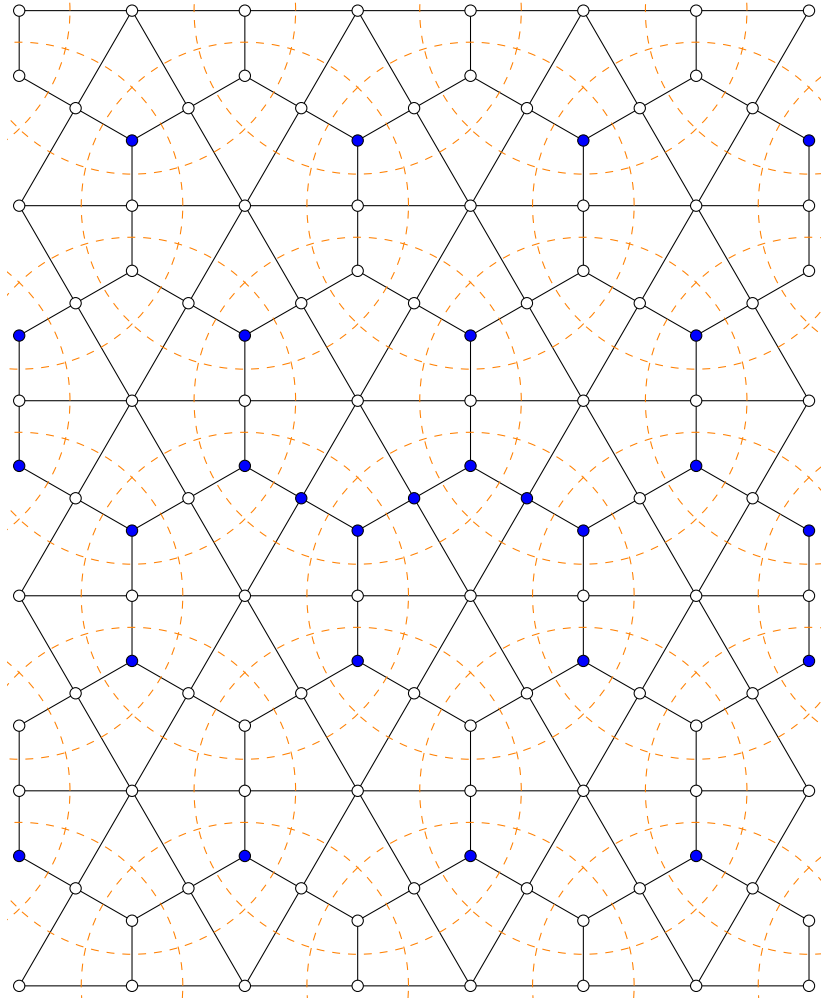
\captionof{figure}{\label{dtt} A well-posed IVP for integrable equations.}}
\end{center}

\section*{Acknowledgements}
It is a pleasure to acknowledge helpful conversations with Yuri Suris, Alexander Bobenko, and James Atkinson during the SIDE 10 conference,
Xikou, Ningbo, China. Many thanks to Vsevolod Adler, who was so kind to translate the relevant sections of his second thesis, where the
content of the original result in \cite{AV} is pointed out clearly. This research has been supported by the Australian Research Council.

\end{document}